\documentclass[11pt,preprint]{aastex}
\pdfoutput=1
\usepackage{graphicx}
\usepackage{amsmath}    % Advanced maths commands
\usepackage{amssymb}    % Extra maths symbols
\let\oldfootsep=\footnotesep
\newcommand\ltsima{$\; \buildrel <\over\sim \;$}
\newcommand\simlt{\lower.5ex\hbox{\ltsima}}
\newcommand\gtsima{$\; \buildrel >\over\sim \;$}
\newcommand\simgt{\lower.5ex\hbox{\gtsima}}

\newcommand\msun {M_\odot}

\newcommand\mearth {{M_\oplus}}

\shorttitle{}
\shortauthors{Bennett et al}

\begin{document}


\title{A Gas Giant Planet in the OGLE-2006-BLG-284L Stellar Binary System}


\author{David P.~Bennett\altaffilmark{1,2,M},
Andrzej~Udalski$^{3,O}$,
Ian A.~Bond\altaffilmark{4,M}, \\
and \\
Fumio Abe$^{5}$, 
Richard K.~Barry$^{1}$, 
Aparna Bhattacharya\altaffilmark{1,2},
Martin Donachie$^{6}$,
Hirosane Fujii$^{7}$,
Akihiko Fukui$^{8,9}$, 
Yuki Hirao$^{1,2,7}$, 
Yoshitaka Itow$^{5}$,  
Kohei Kawasaki$^{7}$,
Rintaro Kirikawa$^{7}$,
Iona Kondo$^{7}$,
Naoki Koshimoto$^{1,2,9,10}$,
Man Cheung Alex Li$^{6}$,
Yutaka Matsubara$^{5}$, 
Shota Miyazaki$^{7}$,
Yasushi Muraki$^{5}$, 
Cl\'ement Ranc$^{1}$,
Nicholas J.~Rattenbury$^{6}$, 
Yuki Satoh$^{7}$, 
Hikaru Shoji$^{7}$,
Takahiro Sumi$^{7}$,
Daisuke Suzuki$^{7}$,
Yuzuru Tanaka$^{7}$,
Paul~J.~Tristram$^{13}$,
Tsubasa Yamawaki$^{7}$,
Atsunori Yonehara$^{6}$,
 \\ (The MOA Collaboration)\\
 Przemek Mr{\'o}z$^{13,3}$,
Radek Poleski$^3$,
Micha{\l} K. Szyma{\'n}ski$^3$,
Igor Soszy{\'n}ski$^3$,
{\L}ukasz Wyrzykowski$^3$,
Krzysztof Ulaczyk$^{14,1}$
 \\ (The OGLE Collaboration)
 } 
              

\affil{$^{1}$Code 667, NASA Goddard Space Flight Center, Greenbelt, MD 20771, USA;    \\ Email: {\tt bennettd@umd.edu}}
\affil{$^{2}$Department of Astronomy, University of Maryland, College Park, MD 20742, USA}
\affil{$^{3}$Astronomical Observatory, University of Warsaw Al.~Ujazdowskie~4,00-478~Warszawa, Poland}
\affil{$^{4}$School of Natural and Computational Sciences, Massey University, Auckland 0745, New Zealand}
\affil{$^{5}$Institute for Space-Earth Environmental Research, Nagoya University, Nagoya 464-8601, Japan}
\affil{$^{6}$Department of Physics, University of Auckland, Private Bag 92019, Auckland, New Zealand}
\affil{$^{7}$Department of Earth and Space Science, Graduate School of Science, Osaka University, Toyonaka, Osaka 560-0043, Japan}
\affil{$^{8}$Department of Earth and Planetary Science, Graduate School of Science, The University of Tokyo, 7-3-1 Hongo, Bunkyo-ku, Tokyo 113-0033, Japan}
\affil{$^{9}$Department of Astronomy, Graduate School of Science, The University of Tokyo, 7-3-1 Hongo, Bunkyo-ku, Tokyo 113-0033, Japan}
\affil{$^{10}$National Astronomical Observatory of Japan, 2-21-1 Osawa, Mitaka, Tokyo 181-8588, Japan}
\affil{$^{11}$Nagano National College of Technology, Nagano 381-8550, Japan}
\affil{$^{12}$University of Canterbury Mt.\ John Observatory, P.O. Box 56, Lake Tekapo 8770, New Zealand}
\affil{$^{13}$Division of Physics, Mathematics, and Astronomy, California Institute of Technology, Pasadena, CA 91125, USA}
\affil{$^{14}$Department of Physics, University of Warwick, Gibbet Hill Road, Coventry, CV4~7AL,~UK}
\affil{$^{M}$MOA Collaboration}
\affil{$^{O}$OGLE Collaboration}

%\clearpage


\begin{abstract}
We present the analysis of microlensing event OGLE-2006-BLG-284, which has a lens system
that consists of two stars and a gas giant planet with a mass ratio of
$q_p = (1.26\pm 0.19) \times 10^{-3}$ to the primary. The mass ratio of the two stars is $q_s = 0.289\pm 0.011$,
and their projected separation is $s_s = 2.1\pm 0.7\,$AU, while the projected separation of the planet from the
primary is $s_p = 2.2\pm 0.8\,$AU. For this lens system to have stable orbits, the three-dimensional
separation of either the primary and secondary stars or the planet and primary star must be much 
larger than that these projected separations. Since we do not know which is the case, the system could 
include either a circumbinary or a circumstellar planet. Because there is no measurement of the
microlensing parallax effect or lens system brightness, we can only make a rough Bayesian
estimate of the lens system masses and brightness. We find host star and planet masses of 
$M_{L1} = 0.35^{+0.30}_{-0.20}\,\msun$, $M_{L2} = 0.10^{+0.09}_{-0.06}\,\msun$, and
$m_p = 144^{+126}_{-82}\,\mearth$, and the $K$-band magnitude of the combined brightness of the
host stars is $K_L = 19.7^{+0.7}_{-1.0}$. The separation between the lens and source system will be
$\sim 90\,$mas in mid-2020, so it should be possible to detect the host system with follow-up adaptive
optics or {\sl Hubble Space Telescope} observations. 
%Follow-up radial velocity observations of the
%binary star system may be able to distinguish the circumbinary and circumstellar possibilities.
\end{abstract}

%\clearpage

\section{Introduction}
\label{sec-intro}

Gravitational Microlensing differs from other exoplanet detection methods do to its sensitivity to low-mass
planets \citep{bennett96} orbiting beyond the snow line \citep{gouldloeb92}, where planet formation is 
thought to be most efficient \citep{lissauer_araa,pollack96}. This unique sensitivity
allows microlensing to yield unique insights into the demographics of these wider orbit planets.
\citet{suzuki16} found a break and likely peak in the mass ratio function that was later 
confirmed to be a peak at a mass ratio of $q_{\rm peak} \simeq 6\times 10^{-5}$ \citep{udalski18,hwang19},
which is close to the Neptune-Sun mass ratio. The smooth, power-law distribution of the \citet{suzuki16,suzuki18}
30-planet sample also does not match predictions of a sub-Saturn mass desert \citep{idalin04} in the exoplanet 
mass ratio distribution. This is thought to be caused by the runaway gas accretion process, which
predicts rapid growth through mass ratios between $10^{-4} \leq q \leq 4\times 10^{-4}$, so that few
planets are expected at these mass ratios. However, this prediction contradicts the \citet{suzuki16,suzuki18}
results, and a comparison to population synthesis models \citep{idalin04,mor09} shows that
these models under-predict the abundance of $10^{-4} \leq q \leq 4\times 10^{-4}$ planets
by a factor of ten, or more if standard prescriptions for planet migration are used. This conclusion that
runaway gas accretion does produce a sub-Saturn mass gap in the planet distribution beyond the snow
line is supported by ALMA observations \citep{nayaskshin19}.

Another region of exoplanet parameter space where microlensing may have unique sensitivity is
for planets in stellar binary systems with star-star or planet-star separations of a few AU
\citep{gouldloeb92}. This is the
separation region where microlensing is most sensitive because it corresponds to the typical Einstein
radius for microlensing events towards the Galactic bulge. Kepler has found a number of circumbinary
planets in closer orbits around relatively tight binaries \citep{doyle11,welch12,welch15,orosz12,kostov13,kostov14},
and a number of these are close to the stability limit where the planetary orbit would become unstable
\citep{holman99}. However, the preponderance of such planets is thought to be a selection effect, as
short period planets are much easier to detect with the transit method. Circumbinary planets in 
wider orbits, like the first
circumbinary planet found by microlensing \citep{bennett16} and the widest orbit circumbinary planet
found by Kepler \citep{kostov16}, are thought to form more easily and be more common than
circumstellar planets \citep{thebault_highig15}.
The short period circumbinary planets are generally thought to have formed in wider orbits and then 
migrated inward to their present positions. (Note that some previous discussion of planets in binary 
systems has used an unfortunate, confusing terminology, referring to circumbinary planets as
``P-type" and circumstellar planets as ``S-type". We reject this nomenclature as unnecessarily confusing,
and we urge other authors to do the same.)

The situation is somewhat different for circumstellar planets in binary systems. The majority of these
systems have very wide stellar binary separations or $> 100\,$AU
\citep{mugrauer09,roell12}, and these wide binary companions are thought to have little effect on planet
formation \citep{thebault_highig15} or stability \citep{holman99}, except in cases where the eccentricity 
of the wide binary pair becomes unstable \citep{kaib13,smullen_kratter16}. 
The situation is different for binary systems with much closer
orbits. However, there are a number of planets orbiting one of a pair of stars in much closer orbits. In particular,
$\gamma$ Cephei A \citep{hatzes03,neuhauser07}, HD 41004 A \citep{zucker04} and HD196885 A
\citep{correia08,chauvin11} are in binary systems with separations of $\sim 20\,$AU and host planets
with $1.6 \,M_{\rm Jup} \simlt M_{Ab} \simlt 2.6 \,M_{\rm Jup}$ with planetary semi-major axes of 
$2.0\,{\rm AU} \simlt a_{Ab} \simlt 2.6\,{\rm AU}$. 

Two similar systems have been found by microlensing, but 
they host much lower mass planets. OGLE-2008-BLG-092LA hosts a planet 
with a mass ratio of $q_{Ab} = 2.4\times 10^{-4}$ with a stellar companion
of mass ratio $q_B = 0.22$ \citep{ogle092}. In units of the Einstein radius, the primary-planet separation 
is $s_{Ab} = 5.26$,
and the primary-secondary separation is $s_{AB} = 17.0$. The masses of the lens system are not known,
but a Bayesian analysis, assuming all lens stars have an equal chance of hosting the observed planet
gives rough estimates of $M_A \sim 0.7\,\msun$ for the host, $M_B \sim 0.15\,\msun$ for the companion,
and $m_{Ab} \sim 57\mearth = 0.18 M_{\rm Jup}$ for the planet. The estimated physical separations
are $a_{Ab} \sim 18\,$AU for the planet and $a_{AB} \sim 58\,$AU, but these are based on the measured
projected separations on the plane of the sky, so one of the separations could be significantly larger than this.
OGLE-2013-BLG-0341LB hosts a planet with a much smaller mass ratio of
$q_p \approx 5\times 10^{-5}$, and the discovery paper \citep{gould14} reports
a microlensing parallax signal that yields measured masses. 
%There is some concern that
%poorly understood systematic errors might lead to a misleading parallax signal, but we will not
%explore this possibility here. 
The masses of the primary, secondary (and planet host) are
determined to be $M_A \approx 0.15 \msun$, $M_B \approx 0.13\msun$ and $m_{Bb} = 2\mearth$.
The projected separations are $a_{AB} \approx 12\,$AU for the two stars and $a_{Bb} \approx 0.8\,$AU
for the secondary and the planet.

As in the case of circumbinary systems, it is thought that the presence of a binary companion can
interfere with planet formation at several different stages \citep{thebault_highig15}. First, binary 
companions can truncate the protoplanetary disk. The inner disk is truncated for circumbinary
planets and the outer disk is truncated for circumstellar planets. This can limit the amount of material
that can be made into planets. Then, the binary companion can also heat up the disk, and this
can interfere with two stages of the core accretion process. First, the initial growth of small grains can
be slowed or halted if the grains collide at high velocities, and second, the planetesimal accumulation
phase can also be slowed or halted by high relative velocities of these km-sized bodies. Thus, it is
thought that planets can only form through the standard core accretion process only in regions that are
not very close to the planetary orbit stability limits found by \citet{holman99}. Thus, the numerous
circumbinary planets found in the Kepler data near this stability limit are thought to have formed
in wider orbits and then migrated inward. Conversely, the five circumstellar planets mentioned above
could presumably have formed in closer orbits and migrated outward. However, outward migration
is thought to be more difficult to achieve than inward migration. Theoretical arguments 
\citep{nelson00,zsom11,picogna13} suggest that a binary companion at 30-50\,AU could inhibit planet
formation through the core accretion method at the grain growth phase, and \citet{paar08} and
\citet{thebault11} argue that the orbits where the  $\gamma$ Cephei Ab and HD196885 Ab planets 
are now located are probably too perturbed to allow the planetesimal accretion process to occur. A search for
binary companions to Kepler planet-hosting stars indicates that stars hosting Kepler planets are 
more than two times less likely have a stellar companion at $< 50\,$AU than stars without a detected
Kepler planet \citep{kraus16}. Thus, the
three planets in binary systems discovered by radial velocities (Cephei Ab, HD196885 Ab, and HD 41004 Ab)
and two planets in binary systems discovered by microlensing 
(OGLE-2008-BLG-092LAb and OGLE-2013-BLG-0341LBb) are expected to have formed in a way more
complicated than the standard core accretion scenario. It could be that the planet or stellar companion
have moved from an orbit that provided a larger separation between the planet and the companion to the
host star, or it could be that the formation process from these planets differs from the standard core
accretion scenario.

In order to understand how such systems form, it would be useful to have more examples. Fortunately,
there are reasons to think that there are additional examples of such systems in existing microlensing
data. \citet{gould14} argued that the discovery of the OGLE-2013-BLG-0341LBb planet was lucky in
the sense that there were two planetary signals, one due to the planetary caustic and one due to the 
central caustic. The planetary caustic signal was very easy to interpret, but the central caustic signal also
implied the presence of the planet, although it was not so easy to identify the planetary signal due to the
central caustic. Because of this, \citet{gould14} argued that there were likely many planetary signals in stellar 
binary events that had yet to be recognized. 
OGLE-2006-BLG-284 is one such event. It is unique in that the
projected separation of the primary star and planet is nearly equal to the projected separation between the primary
and secondary stars. We should note, however, that there is one other event published as 
planet in a binary system, with very similar planet-primary and secondary-primary separations, 
OGLE-2016-BLG-0613 \citep{han_ob160613}, but unpublished MOA data appears to
contradict the published models\footnote{See the video from the 2019 Microlensing Conference
day 3, part 3, at about the 30 minute mark, available at
\tt https://www.simonsfoundation.org/event/23rd-international-microlensing-conference/}. 
An acceptable model has not yet been found for this event, but 
the photometry data will be provided by the first author upon request.

This paper is organized as follows. Section~\ref{sec-lc_data} discusses the data set and photometry,
and Section~\ref{sec-lc} presents the light curve modeling. We present an
extinction estimate, and the source angular radius in Section~\ref{sec-radius}. In Section~\ref{sec-lens_prop},
we derive the properties of the lens system that can be determined from the light curve analysis, and
finally, in Section~\ref{sec-conclude}, we discuss how future observations may be able to improve our
understanding of this system planets in binary systems in general.
 
\section{Light Curve Data and Photometry}
\label{sec-lc_data}

Microlensing event OGLE-2006-BLG-284, at ${\rm RA} =17$:58:38.22, 
${\rm  DEC} = -29$:08:12.0, and Galactic coordinates $(l, b) = (1.2771, -2.5505)$, was 
identified and announced as a microlensing candidate by the Optical Gravitational Lensing 
Experiment (OGLE) Collaboration as a part of the OGLE-III survey \citep{udalski08}.
It was identified as a binary microlensing event in an OGLE catalog of binary events from
2006-2008 \citep{jar10}, but this paper contained no discussion of the feature due to the planet
that does not fit the binary microlensing model.
This event was discovered in the OGLE-III bulge field BLG206, but it occurred a region of sky where two
bulge fields overlap, and so there is data from OGLE-III bulge field BLG205 for this event, as well.
The event was not detected by the MOA alert system, which was only partly functional in
2006, due to the lack of baseline data available in the first year of the MOA II survey.
The event was found in the 9-year (2006-2014) retrospective analysis of the MOA II survey
data, which included systematic modeling of all binary lens events that has yielded a
number of newly discovered planetary events \citep{kondo19}. This data is now available
at the NASA Exoplanet archive\footnote{\tt https://exoplanetarchive.ipac.caltech.edu/docs/MOAMission.html}
under the star ID gb9-R-3-6-14546. While this public data set leads to the same conclusions, we have
used a modified version of the MOA difference imaging pipeline \citep{bond01,bond17} that 
automatically calibrates the photometry to the OGLE-III catalog \citep{ogle3-phot}.
Both the MOA and OGLE 
photometry used the difference imaging method \citep{tom96,ala98}.
The OGLE Collaboration provided 
optimal centroid photometry using the 
OGLE difference imaging pipeline \citep{ogle-pipeline}. 

\section{Light Curve Models}
\label{sec-lc}

Our light curve modeling was done using the image centered ray-shooting method
\citep{bennett96} with the initial condition grid search method described in
\citet{bennett-himag}. As is typical for triple lens events, the modeling took place in two stages.
First, we searched for stellar binary solutions to the OGLE and MOA data sets with the 
observations with $3892.2 < t < 3895.3$ removed. (We define our time parameter to be
the modified heliocentric Julian Day, ${\rm HJD} - 2450000$.) This search used the
modified version of the initial condition grid search method that replaced the 
Einstein radius crossing time, $t_E$, and time of closest approach between the lens center-of-mass 
and the source star, $t_0$, with the times of the caustic entry and exit. This method greatly speeds
up the search for the best solutions. This search led to a unique best fit solution, with the next best 
binary lens model
disfavored by $\Delta\chi^2 = 372.6$.

\begin{figure}
\epsscale{0.9}
\plotone{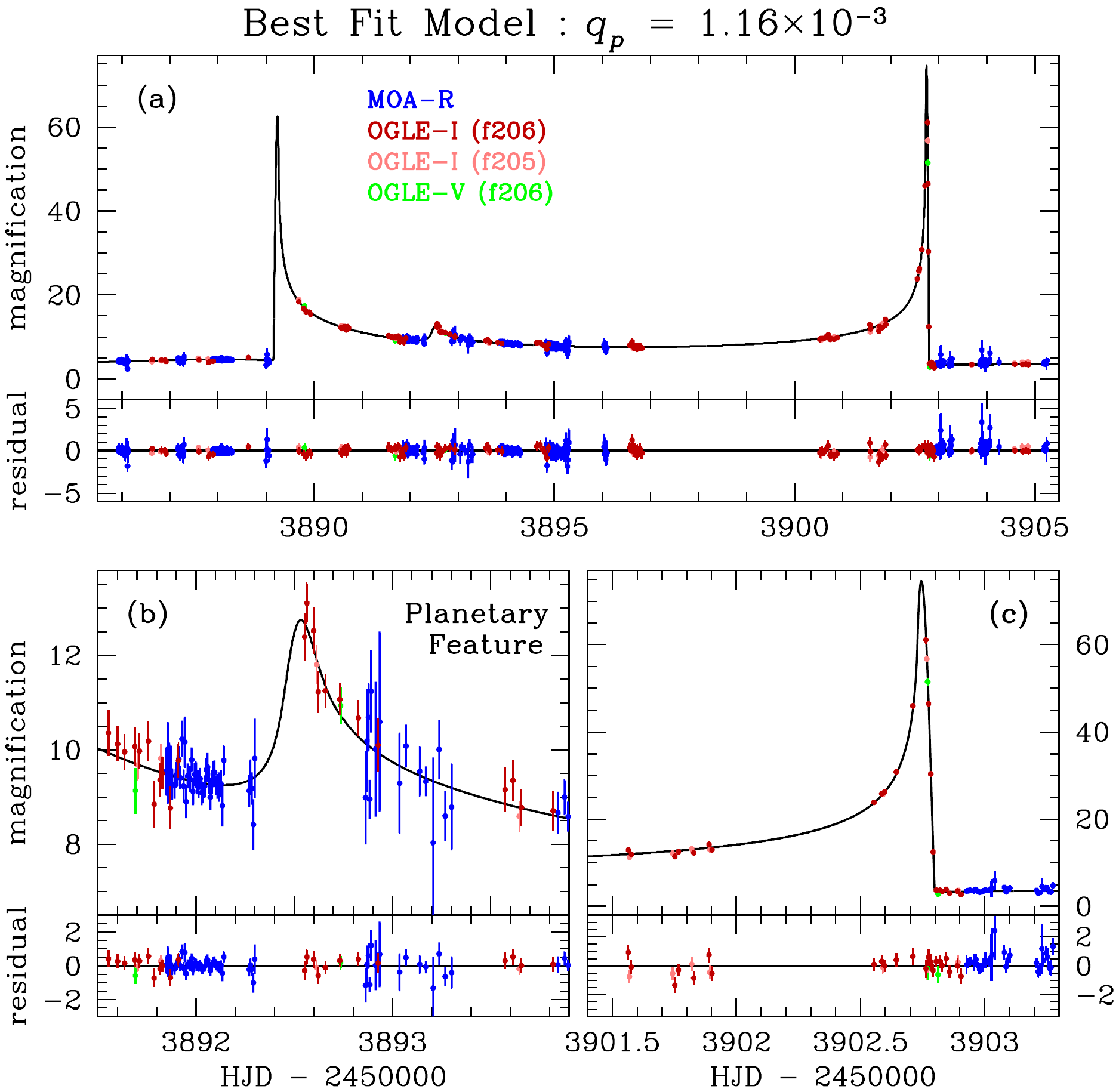}
\caption{The best triple lens model for  the OGLE-2006-BLG-284 light curve.
The MOA-red data are shown in blue while the OGLE $I$-band data from fields BLG206
and BLG205 are shown in dark and light red, respectively. The OGLE $V$-band data from field
BLG206, is shown in green. The solid line is the best fit
model. Panel (a) shows the magnified region of the stellar binary light curve, and
panels (b) and (c) show close-ups of the planetary feature and the binary caustic exit resolved
by OGLE that enable the measurement of the source radius crossing time, $t_*$.
\label{fig-lc}}
\end{figure}

\begin{figure}
\epsscale{1.0}
\plotone{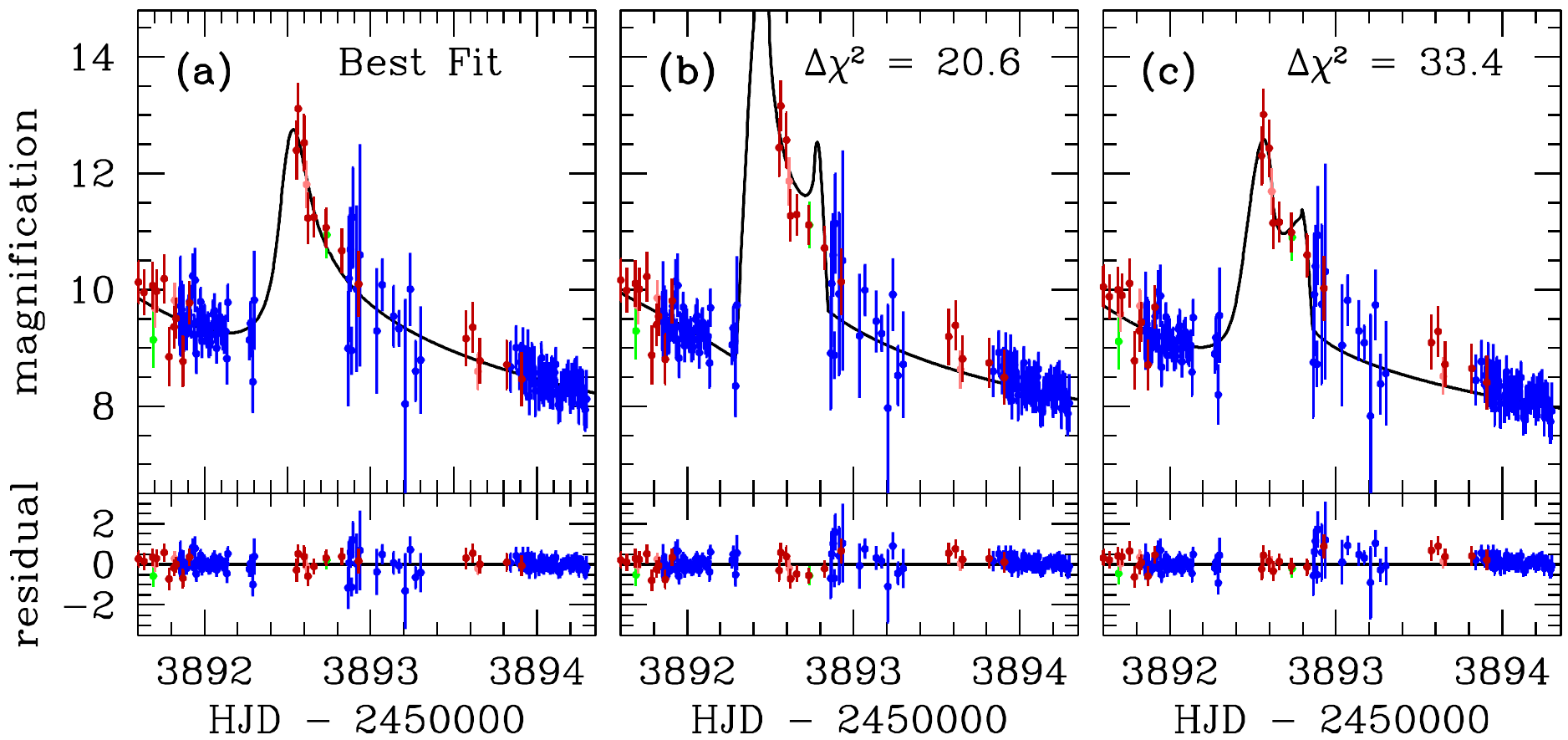}
\caption{Comparison of the planetary features for the 3 best 2-star plus 1-planet
models. The planetary feature in the best fit model, which has a planetary cusp
approach feature is shown in panel (a). Panels (b) and (c) show the second and
third best models which feature planetary caustic crossings. The parameters of
these models are given in Table~\ref{tab-mparams}.
\label{fig-lc_comp}}
\end{figure}

The second step was to fix the binary lens parameters and search for a triple lens model
that could account for the feature at $3892 < t < 3893.5$ in the light curve (see Figure~\ref{fig-lc}).
(We define our time parameter as the modified Heliocentric Julian Day, $t = {\rm HJD} -2450000$.)
This analysis led to a number of possible solutions that maintained basically the same light
curve away from this short duration anomaly. Close-ups of this short duration light curve anomaly 
for the three best triple lens light curve solutions 
are shown in Figure~\ref{fig-lc_comp}.

\begin{figure}
%\epsscale{0.7}
\plottwo{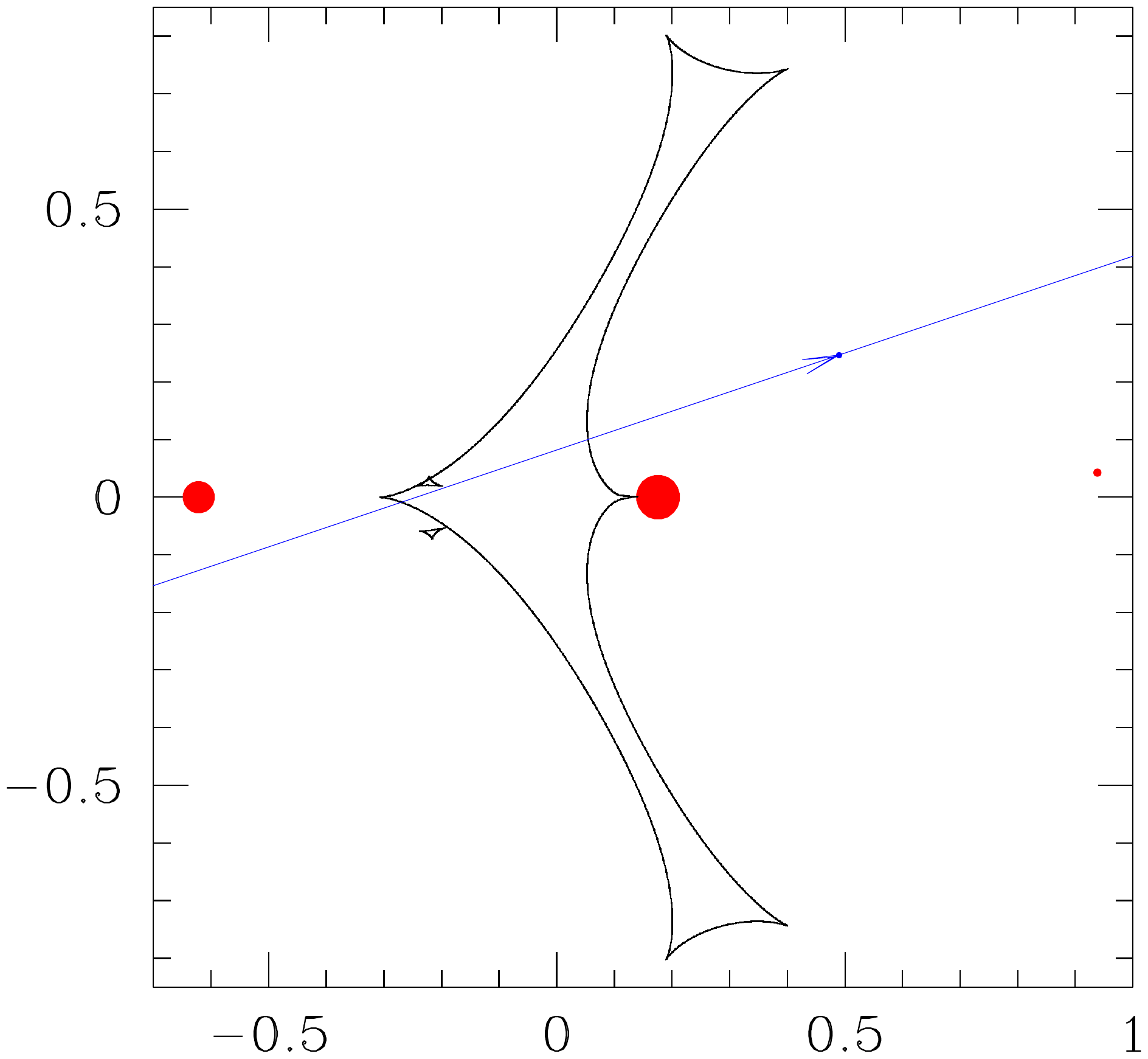}{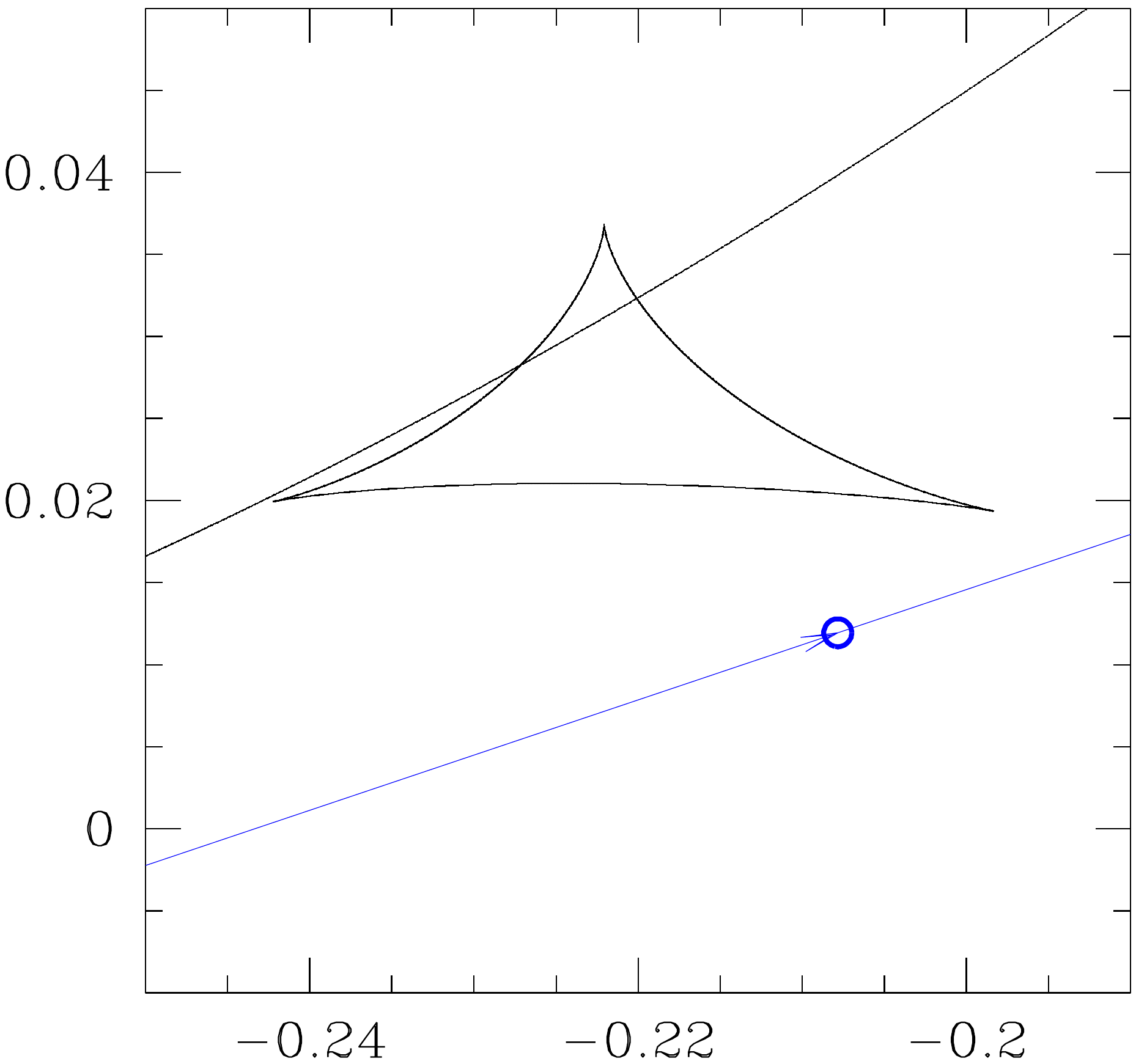}
\caption{The caustic configuration for the best fit model is plotted in units of the 
Einstein radius. The line with the arrow represents the motion of the center of the source
star, and the blue line and circle indicates the trajectory and
size of the source star. The red dots in the right panel
indicate the positions of the lens stars and planet. The close-up of the top planetary 
caustic in the right panel indicates that the planetary caustic overlaps with the stellar
binary caustic, but has no clear interaction with it due to the fact that the caustics are
related to distinct, different images.
\label{fig-caustic}}
\end{figure}

The best fit model, shown in Figure~\ref{fig-lc}, has a planetary cusp approach at
${\rm HJD} -2450000 = t = 3892.5$. The caustic structure for this lens system is
shown in Figure~\ref{fig-caustic}, and it is somewhat unusual for a triple lens system because
the planetary caustic crosses the stellar binary caustic with no interaction with it,
although \citet{danek15,danek19} have shown some examples like this. This is because 
the caustics affect different images. (There are 4 images when the source is outside all the
caustics, 6 images when the source is inside one caustic, and 8 images when it is inside
two caustic curves.) The parameters of the best fit model shown in Figure~\ref{fig-lc}, as
well as the two best planetary caustic crossing models, are given in Table~\ref{tab-mparams}.
The parameters that these model
has in common with a single lens model are the Einstein radius crossing time, $t_E$, 
and the time, $t_0$, and distance, $u_0$, of closest approach between the lens center-of-mass 
and the source star. We use the triple lens parameter system of the first published triple lens
system \citep{gaudi-ogle109,bennett-ogle109}, with one minor modification. We use the mass
ratios to the primary star instead of mass fractions of the total lens system mass.
For this event, we assign the primary star to be mass 3, the secondary star to be mass 1, and the planet to be
mass 2. The two mass ratio parameters are then $q_1$ for the secondary star, and $q_2$ for the planet.
The separation between masses 2 and 3 is given by $s_{23}$ and the separation between mass 1
and the center of mass for masses 2 and 3 is given by $s_{1\rm cm}$. The angle between the source
trajectory and the axis connecting mass 1 with the center-of-mass for masses 2 and 3 is given by
$\alpha_{1\rm cm}$, and the angle between this axis and the line connecting masses 2 and 3 is
given by $\phi_{23}$. The final lens model parameters are the source radius crossing time, $t_*$, 
and the $I$ and $V$ band source magnitudes, $I_S$ and $V_S$.
The length parameters, $u_0$, $s_{1\rm cm}$ and $s_{23}$, are normalized by the Einstein radius of the 
total lens system mass, 
$R_E = \sqrt{(4GM/c^2)D_Sx(1-x)}$, where $x = D_L/D_S$ and $D_L$ and $D_S$ are
the lens and source distances, respectively. ($G$ and $c$ are the Gravitational constant
and speed of light, as usual.) 

\begin{deluxetable}{cccccc}
\tablecaption{Best Model Parameters
                         \label{tab-mparams} }
\tablewidth{0pt}
\tablehead{
%% Use a footnote to explain numbering.
%& & & & \multicolumn{2}{c} {MCMC averages} \\
\colhead{parameter}  & \colhead{units} &
\colhead{best fit} & \colhead{caustic-1} & \colhead{caustic-2} & \colhead{MCMC averages}
}  % end header.

\startdata

$t_E$ & days &39.677 & 39.872 & 39.969 & $39.38 \pm 0.97$  \\
$t_0$ & ${\rm HJD}-2453800$ & 99.4440 & 99.4060 & 99.2922 & $99.425\pm 0.066$  \\
$u_0$ & & 0.077523 & 0.079517 & 0.081596 & $0.0776\pm 0.0020$  \\
$s_{1\rm cm}$ & & 0.79801 & 0.79856 & 0.79881 & $0.8007 \pm 0.0091$  \\
$s_{23}$ & & 0.76396 & 0.78233 & 0.78404 & $0.7615\pm 0.0063$ \\
$\alpha_{1\rm cm}$ & radians & 0.32432 & 0.33979 & 0.35519 & $0.3237\pm 0.0039$  \\
$\phi_{23}$  & radians & 0.05590 & 0.02075 & -0.02990 & $0.0607\pm 0.0093$ \\
$q_1$ &  & 0.28445 & 0.27180 & 0.27395 & $0.289 \pm 0.011$  \\
$q_2$ & $10^{-4}$ & 11.632 & 4.2933 & 0.71234& $12.6 \pm 1.9$  \\
$t_\ast$ & days & 0.03367 & 0.03321 & 0.03311 & $0.03372\pm 0.00030$ \\
$I_s$ & & 20.007 & 20.012 & 19.998 & $19.995\pm 0.036$  \\
$V_s$ & & 21.815 & 21.827 & 21.807 & $21.803\pm 0.036$  \\
fit $\chi^2$ &  & 12414.60 & 12435.16 & 12447.96 & \\
for 12407 & d.o.f. & & & & \\

\enddata
\end{deluxetable}

For every passband, there are two parameters to describe the unlensed source
brightness and the combined brightness of any unlensed ``blend" stars that are
superimposed on the source. Such ``blend" stars are quite common because
microlensing is only seen if the lens-source alignment is $\simlt \theta_E \simlt 1\,$mas,
while stars are unresolved in ground based images if their separation is
$\simlt 1^{\prime\prime}$. These
source and blend fluxes are treated differently from the other parameters because
the observed brightness has a linear dependence on them, so for each set of 
nonlinear parameters, we can find the source and blend fluxes that minimize the
$\chi^2$ exactly, using standard linear algebra methods \citep{rhie_98smc1}.

The MOA data immediately after the planetary light curve feature at $t \simeq 3892.6$ are crucial
for excluding caustic crossing planetary models that would have a caustic exit at $3893.0 < t < 3893.5$,
and the best remaining planetary caustic crossing models are the second and third best models 
with planetary features shown in Figures~ \ref{fig-lc_comp}b and c.
The $\chi^2$ improvement for the best fit model, shown in Figures~\ref{fig-lc} and \ref{fig-lc_comp}a,
over these competing models
is relatively small, with $\chi^2$ improvements of $\Delta\chi^2 = 20.6$, and $\Delta\chi^2 = 33.4$ over
the second and third best models (shown in Figure~\ref{fig-lc_comp}b and c), respectively. While these
$\chi^2$ improvements are rather modest, we believe that they are sufficient to exclude these second
and third best models. The best fit model is preferred by both the OGLE and MOA data sets.
The $\chi^2$ difference between the second best fit model and the best fit model is occurs primarily at
$3892 < t < 3895$, while the $\chi^2$ difference between the third best and best fit models occurs primarily at
$3893 < t < 3896$. Both the second and third best models have planetary caustic exit features that 
the data has no indication. For the second best model (Figure \ref{fig-lc_comp}b), the caustic exit
is mostly squeezed between two data points, but it does get a modest $\chi^2$ penalty because 
the caustic exit feature is a bit wider than the gap between the data points at $t = 3893.73$ and
$t = 3893.83$. Similarly, this model also has strong caustic entrance just after the last MOA data point
of the previous night at  $t = 3892.299$. The timing of these two caustic features requires rather unlikely
coincidences, so this lends credence to the idea that the best fit model is the correct one.  The
third best model (Figure \ref{fig-lc_comp}c) has such a weak planetary caustic exit that it does not 
get a significant $\chi^2$ penalty, although the data also provide no caustic exit signal. Both the
second and third best models predict a lower magnification than is observed in the time after these planetary 
caustic exits, $3892.86 < t < 3895$, and the data in this range contribute substantially
to the $\chi^2$ differences between these two models and the best fit model. Because these
competing models have additional circumstantial evidence against them, we are comfortable in treating their
likelihood with the Gaussian probability, $e^{-\Delta\chi^2/2}$, in the analysis that follows.

In addition to triple lens models, we also searched for binary source, binary lens models, sometimes
referred to as 2L2S models. The best 2L2S model we found has a $\chi^2$ larger than
our best triple lens (3L1S) model by $\Delta\chi^2 = 552.56$. This is not surprising because it would
require a much fainter second source to explain the low amplitude feature that we attribute to the
planet as a feature due to the stellar binary. However, if the second source is much fainter than the 
primary, then it should also be much redder, but a much redder source would cause a shift in the 
MOA-red data and a large shift in the OGLE $V$ band data with respect to the OGLE $I$ band data
at the time of the feature that we attribute to the planet. Therefore, we will not consider these models
further.

\section{Photometric Calibration and Source Radius}
\label{sec-radius}

The light curve models listed in Table~\ref{tab-mparams} constrain the finite source
size through measurement of the source radius crossing time, $t_*$,  and this allows us to derive the angular Einstein
radius, $\theta_E = \theta_* t_E/t_*$, if we know the angular size of the source star, $\theta_*$.
This can be derived from the extinction corrected brightness and color of source star
\citep{kervella_dwarf,boyajian14}. 

\begin{figure}
\epsscale{0.9}
%\plotone{cmd_mb11291_VIhstrS3.pdf}
\plotone{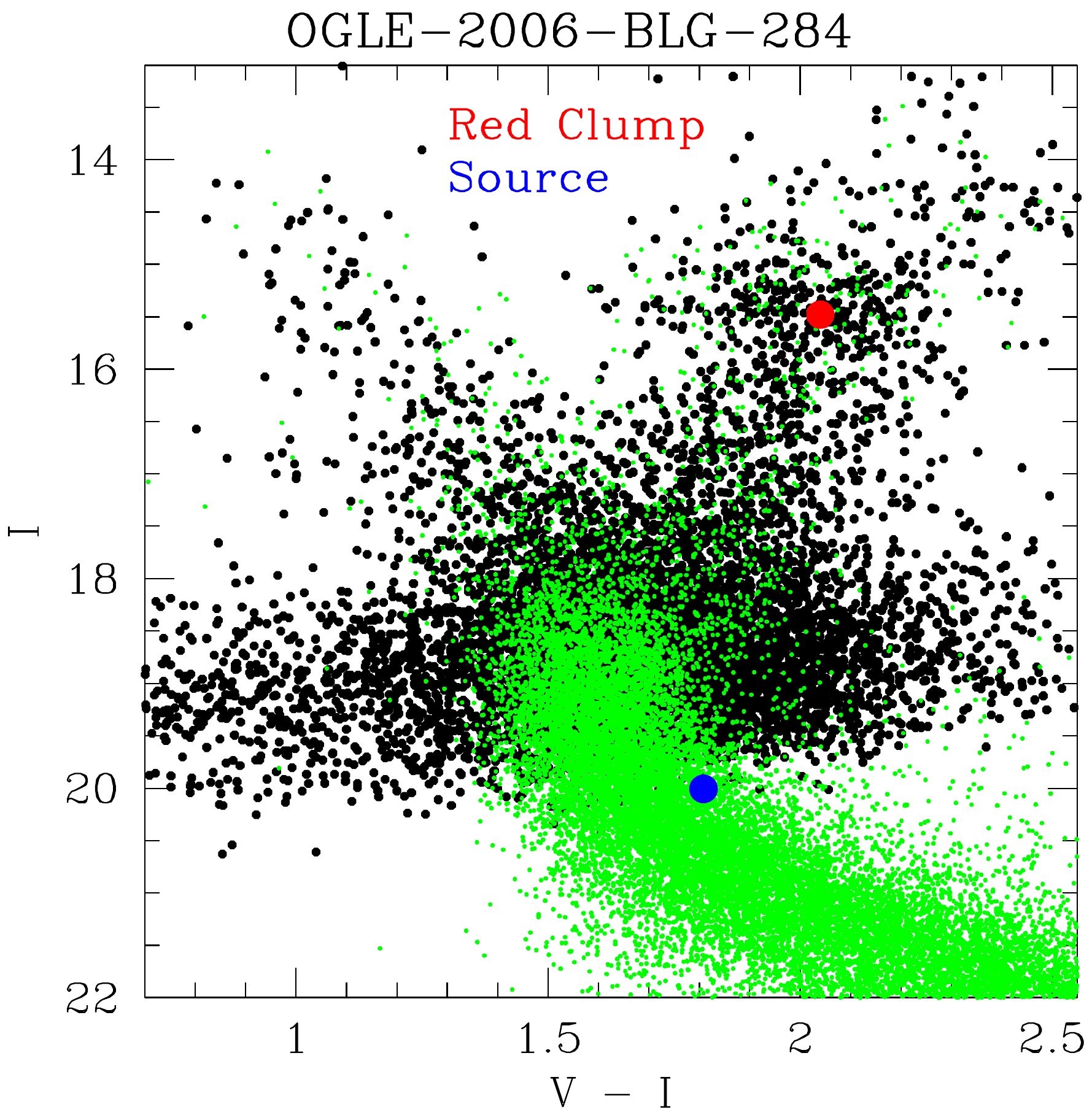}
\caption{The $(V-I,I)$ color magnitude diagram (CMD) of the OGLE-III stars within 
$90^{\prime\prime}$ of OGLE-2006-BLG-284 transformed to calibrated
Johnson $V$ and Cousins $I$ using the transformation given by \citet{ogle3-phot}. The red spot indicates
red clump giant centroid, and the blue indicates the source magnitude
and color. The green dots represent the HST Baade's Window CMD of \citet{holtzman98} transformed
to the extinction and Galactic longitude appropriate for this field.
\label{fig-cmd}}
\end{figure}

In order to estimate the source radius, we need extinction-corrected magnitudes,
which can be determined from the magnitude and color of the centroid of the
red clump giant feature in the CMD \citep{yoo_rad}, as indicated
in Figure~\ref{fig-cmd}. We find that the
red clump centroid in this field is at $I_{\rm cl} = 15.48$, $(V-I)_{\rm cl} = 2.04$, 
which implies $V_{\rm cl} = 17.52$. From \citet{nataf13}, we find that the extinction corrected
red clump centroid should be at $I_{\rm cl,0} = 14.39$, $(V-I)_{\rm cl0,} = 1.06$, which implies
$I$ and extinctions of $A_I = 1.09\pm 0.05$ and a color excess of $E(V-I) = 0.098\pm 0.03$. So, the extinction corrected
source magnitude and color are $I_{s0} = 18.905$ and $(V-I)_{s0} = 0.828$ for the average model
from our MCMC calculations reported in Table~\ref{tab-mparams}.
These dereddened magnitudes can be used to determine the angular source radius,
$\theta_*$. With the source magnitudes that we have measured, the most precise determination
of $\theta_*$ comes from the $(V-I),I$ relation. We use
\begin{equation}
\log_{10}\left[2\theta_*/(1 {\rm \mu as})\right] = 0.501414 + 0.419685\,(V-I)_{s0} -0.2\,I_{s0} \ ,
\label{eq-thetaS}
\end{equation}
which comes from the \citet{boyajian14} analysis, but with the color range optimized for the 
needs of microlensing surveys. These numbers were provided in a private communication from 
T.S.\ Boyajian (2014). Therefore, we handle this 
uncertainty in our MCMC calculations, so as to include all the correlations in our determination
of the lens system properties.
For the average model parameters from our MCMC calculation, listed in Table~\ref{tab-mparams}, we find 
$\theta_* = 0.585\pm 0.029\,\mu$as.

Figure~\ref{fig-cmd} also includes the location of the source star on the CMD, in blue, and the 
{\it Hubble Space Telescope} $V$ and $I$ band CMD from \citet{holtzman98} after it has been 
shifted to the average distance and extinction of the red clump stars in the OGLE-2006-BLG284 field. 
The position of the
source star in Figure~\ref{fig-cmd} indicates that it is on the red edge of the main sequence. However,
the CMD is likely to have a larger dispersion in the field of this event, due to the higher extinction in
this field.

\section{Lens System Properties}
\label{sec-lens_prop}

\begin{deluxetable}{cccc}
\tablecaption{Physical Parameters\label{tab-pparams}}
\tablewidth{0pt}
\tablehead{
\colhead{Parameter}  & \colhead{units} & \colhead{Value} &  \colhead{2-$\sigma$ range} \\
}  % end header.
\startdata 
$\theta_E$ & mas & $0.681\pm 0.035$  & 0.612-0.750  \\
$\mu_{\rm rel,G}$ & mas/yr & $6.31\pm 0.33$ & 5.66-6.97 \\
$D_S $ & kpc & $8.6\pm 1.2$  & 6.2-10.7 \\
$D_L $ & kpc & $4.0\pm 1.5$  & 1.2-6.5 \\
$M_{L1}$ & $\msun$ & $0.35^{+0.30}_{-0.20}$  & 0.06-0.84   \\
$M_{L2}$ & $\msun$ & $0.100^{+0.097}_{-0.056}$  & 0.018-0.24   \\
$m_p$ & $\mearth$ & $144^{+137}_{-82}$  & 25-375  \\
$a_{\perp,ss}$ & AU & $2.06\pm 0.74$  & 0.61-3.33   \\
$a_{\perp,s1p}$ & AU & $2.17\pm 0.78$  & 0.65-3.51   \\
$V_L$ & mag & $26.4^{+1.8}_{-2.4}$ & 22.3-37.2 \\
$I_L$ & mag & $22.9^{+1.1}_{-1.6}$ & 20.2-31.0 \\
$K_L$ & mag & $19.7^{+0.7}_{-1.0}$ & 17.8--26.8 \\
\enddata
\tablecomments{ Mean values and RMS are given for $\theta_E$, $\mu_{\rm rel,G}$, $D_S$, $D_L $, $a_{\perp,ss}$, 
and $a_{\perp,sp}$. 
Median values and 68.3\% confidence intervals are given for the other parameters. 2-$\sigma$ range refers to the
central 95.3\% confidence interval.}
\end{deluxetable}

With our determination of $\theta_*$ from the source magnitude and color in Section~\ref{sec-radius}, we can
now proceed to determine the lens system properties. The angular Einstein radius is given by
$\theta_E = \theta_* t_E/t_*$, which allows us to use the following relation \citep{bennett_rev,gaudi_araa}
\begin{equation}
M_L = {c^2\over 4G} \theta_E^2 {D_S D_L\over D_S - D_L} 
%       =  {c^2\over 4G} \theta_E^2 {{\rm AU}\over \pi_{\rm rel}}
       = 0.9823\,\msun \left({\theta_E\over 1\,{\rm mas}}\right)^2\left({x\over 1-x}\right)
       \left({D_S\over 8\,{\rm kpc}}\right) \ ,
\label{eq-m_thetaE}
\end{equation}
where $x = D_L/D_S$ to determine the relationship between the lens system mass, $M_L$ and
distance, $D_L$. We know that the source is very likely to be approximately at the distance
of the Galactic bulge, but the bulge is bar shaped and pointed at approximately the location of
our Solar system. As a result, there is an uncertainty of a few kpc in the distance to the source
star, $D_S$. Also, the probability of the lens system mass and distance also depend on the 
the Geocentric lens-source relative proper motion, which can be determined from the
angular source size, the source angular radius, and the source radius crossing
time $\mu_{\rm rel,G} = \theta_*/t_*$. We can combine all these factors with a Galactic
model prior to determine our best estimate of the properties of this binary star plus planet
system. We use the Galactic model of \citet{bennett14}, and we assume that the planet
hosting probability is independent of the planet mass, because this is the simplest 
assumption to make. We also include the 2nd and third best models in our collection of
MCMC models to combine with the Galactic prior, but we weight them by $e^{-\Delta\chi^2/2}$, 
which gives them very little weight.

\begin{figure}
\epsscale{1.08}
%\plotone{lens_prop_moa291_a3d_LS.pdf}
\plotone{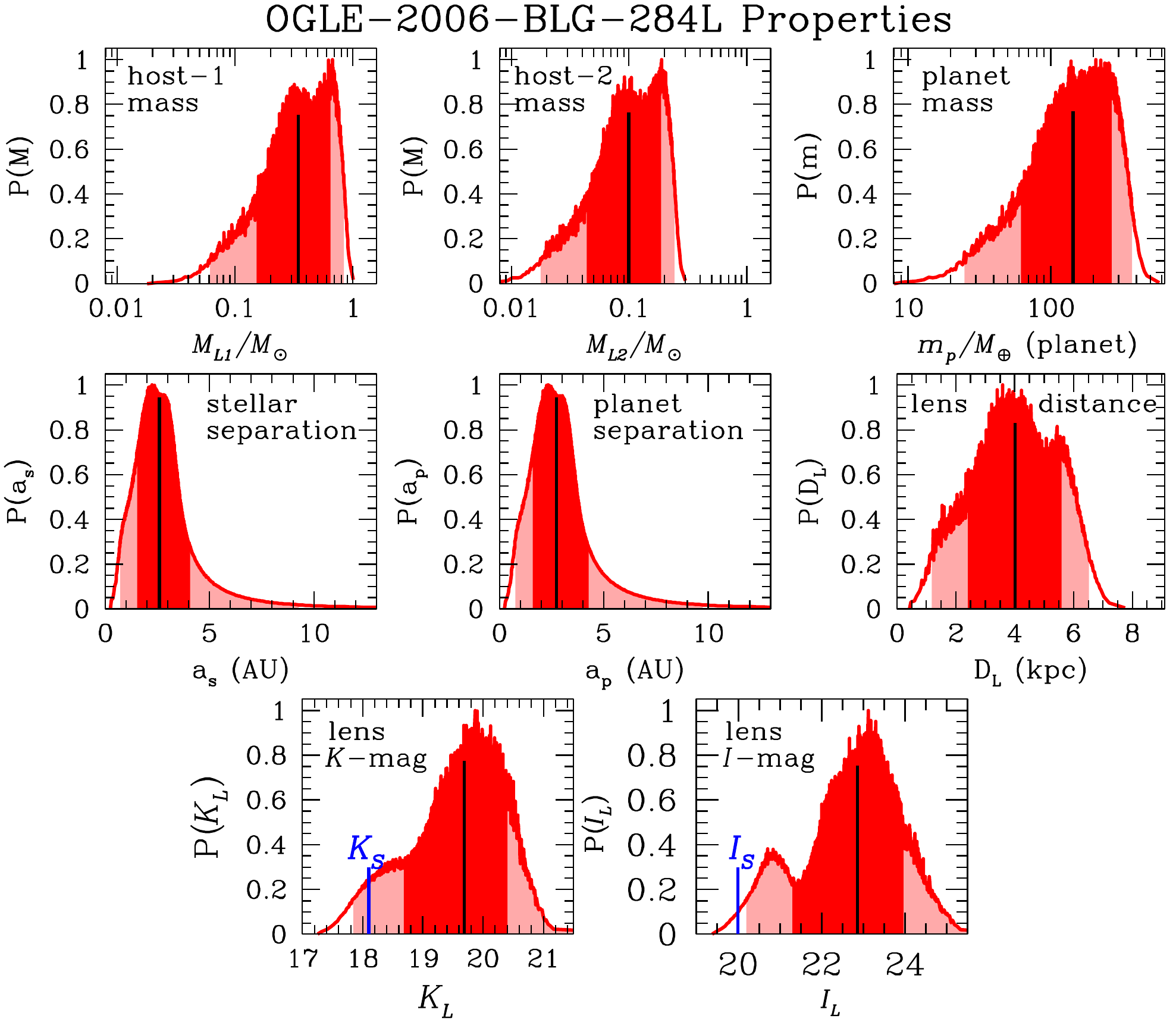}
\caption{The lens properties from our Bayesian analysis. Host-1 and Host-2 are the
primary and secondary stars of the system. The stellar and planetary separations are the
separations from the primary star. The blue lines in the bottom two panels indicate the
source magnitudes.
\label{fig-lens_prop}}
\end{figure}

The results are presented in Figure~\ref{fig-lens_prop} and Table~\ref{tab-pparams}.
This table and figure introduce some new parameters. The primary and secondary stellar masses
are given by $M_{L1}$ and $M_{L2}$, and the planet mass is $m_p$. Table~\ref{tab-pparams} reports
the projected separations, $a_{\perp,ss}$ and $a_{\perp,s1p}$, between the two stars, and between
the primary star and planet, respectively. Instead of these projected separations, Figure~\ref{fig-lens_prop},
shows the predicted distribution of 3-dimensional separations, under the assumption of random
orientations. However, the orientations of the primary-secondary stellar separation and the 
primary star-planet separations cannot be random. They must be anti-correlated in order to obey
orbital stability requirements \citep{holman99}. For a circumstellar planet orbiting the
primary star, with the observed secondary-primary mass ratio of $0.289\pm 0.011$, the 
semi-major axis of the planet must be $\leq 0.38$ times the semi-major axis of the stellar binary system,
if we assume that both orbits are nearly circular. In the circumbinary case, \citet{holman99} find
that the planet must have a semi-major axis $\geq 2.2$ times the semi-major axis of the stellar binary system,
assuming nearly circular orbits. Since the observed separations in Table~\ref{tab-pparams} are nearly 
identical, this implies that the 3-dimensional separation of either the primary star and the planet or the
two planets must be $\geq 2.2$ times the projected separations.

The bottom two panels of Figure~\ref{fig-lens_prop} have blue lines that indicate the magnitudes of the combined 
light from both source stars as vertical blue lines. In June, 2020, it will be 14 years since this event has 
occurred, so the lens-source separation should be $>88.3 \pm 4.6\,$mas after June 2020. Our
experience with Keck adaptive optics observations \citep{bennett20,aparna20,terry20} indicates
that the lens should be detectable at this separation if it is within 3 magnitudes of the source
brightness, and our Bayesian analysis indicates that there is a 96.4\% chance that the lens
system is $\leq 3\,$mag fainter than the source stars.

\section{Conclusions}
\label{sec-conclude}

Our analysis of OGLE-III data from two fields and MOA data from the MOA 9-year
retrospective analysis for microlensing event OGLE-2006-BLG-284 has revealed that
this event is due to a triple lens system consisting of two stars and a planet. While there are 
two light curve models with $\chi^2$ values that are only $\Delta\chi^2 = 20.6$ and 33.4 higher
than the best fit solution, we argue that these also have very unlikely parameters. And we do 
not consider these to be possible solutions for this event.

The prime stumbling block for the interpretation of this event is the fact that the projected separation 
of the primary and secondary stars is nearly identical to the projected separation between the planet
and the primary star. Thus, we cannot tell from the light curve data alone whether planet orbits both stars
(circumbinary) or only one (circumstellar). Fortunately, because the event occurred so long ago, the
lens-source separation is now large enough ($\simgt 88\,$mas) that it should be readily detected by
Keck adaptive optics observations, which has now been done for a number of host stars of
microlens planets 
\citep{gaudi-ogle109,bennett-ogle109,bennett20,batista15,beaulieu16,beaulieu18,aparna18,aparna20,terry20}.
This may provide an opportunity to make observations that can decide between the
circumbinary and circumstellar possibilities with the next generate of large telescopes. 
A time series of radial velocity observations of the lens
system with a precision of $\sim 2\,$km/sec should be able to detect the radial velocity of the 
primary star if it has a relatively small semi-major axis of $\simlt 2\,$AU. This would imply that 
two stars were in the relatively close orbit and that the planet would have to be in a much wider orbit, with
a large line-of-sight separation. Conversely, the failure to detect any radial velocity for the primary lens star at the 
$\sim 2\,$km/sec level would imply that the primary and secondary stars have a wide orbit and the planet
orbits only the primary.

The details of any radial velocity program will depend on the brightness of the lens star system, which
has yet to be determined. The $K_L$ histogram in Figure~\ref{fig-lens_prop} indicates that there is a 
95\% chance that the lens system magnitude is in the range $17.5 < K_L < 20.9$, but this also
represents a range of lens system distances of $1.0\,{\rm kpc} < D_L < 7.5\,$kpc and a primary star
mass range of $0.08\msun < M_{L1} < 1.05\msun$. Since the star-star and planet-star separations, as
well as the orbital period, are proportional to $D_L$, this implies a large range of possible masses and
orbital periods, as well as a 3.4 magnitude brightness range. Thus, we cannot really determine the
observational effort needed to test the circumbinary vs.\ circumstellar interpretations until we have
detected the lens star with high angular resolution imaging. Even in the most optimistic scenario
with $K_L < 18$, current telescopes and instruments are unlikely to be able to make such measurements,
but there is some hope for observations with the next generation of extremely large telescopes (ELTs).
However, this may require advanced instrumentation that may not be available at first light on these
ELTs.

This discovery tends to confirm the \citet{gould14} prediction that there are likely to be many undiscovered
planetary signals in the existing catalog of observed microlensing events, and it may help to reduce the 
deficit of planets in binary systems noted by \citet{koshimoto20}.
Such events probe a range
of planets in binary systems that are difficult to explore with other methods, and so they are likely to
provide a unique new window on the formation of planets in binary systems and the planet formation
process, in general.

The search for planets in binary systems would benefit from more efficient triple-lens modeling methods
and more efficient methods to search the triple lens parameter space. This problem will come more acute 
with the launch of NASA's Wide Field Infrared Survey Telescope (WFIRST) \citep{WFIRST_AFTA},
which will devote a large fraction of its observing time to an exoplanet microlensing survey
\citep{bennett02,bennett18_wfirst,penny19}. This will provide a much larger sample of exoplanets than can be
observed from the ground, and will have much better light curve sampling than any ground based 
survey. A more robust triple-lens modeling system will be needed to take advantage of the 
circumbinary and circumstellar planets that will be detectable in the WFIRST data.

\acknowledgments 
DPB, AB, and CR  were supported by NASA through grant NASA-80NSSC18K0274.
The MOA project is supported in Japan by JSPS KAKENHI Grant Numbers JP17H02871, JSPS24253004, 
JSPS26247023, JSPS23340064, JSPS15H00781, and JP16H06287.
The work by CR was supported by an appointment to the NASA Postdoctoral Program at the 
Goddard Space Flight Center, administered by USRA through a contract with NASA.
NJR is a Royal Society of New Zealand Rutherford Discovery Fellow.
The OGLE Team thanks Profs.\ Marcin Kubiak and Grzegorz Pietrzy{\'n}ski for
their contribution to the OGLE photometric data. The OGLE project has
received funding from the National Science Centre, Poland, grant MAESTRO
2014/14/A/ST9/00121 to AU.


\begin{thebibliography}{}
%\bibitem[Alard(1997)]{alard97} Alard, C.\ 1997, \aap, 321, 424 
\bibitem[Alard \& Lupton (1998)]{ala98}Alard, C. \& Lupton, R.H.\ 1998, \apj, 503, 325
%\bibitem[Albrow et al.(2000)]{albrow-97blg41} Albrow, M.D.\ 2000, \apj, 534, 894
%\bibitem[Albrow et al.(2009)]{albrow09} Albrow, M.~D., Horne, K., Bramich, D.~M., et al.\ 2009, \mnras, 397, 2099
%\bibitem[Alcock et al.(1995)]{macho-par1}Alcock, C., Allsman, R.~A., Alves, D., et al.~1995, \apjl, 454, L125
%\bibitem[Alcock et al.(1997)]{macho-95b30} Alcock, C., Allen, W.H., Allsman, R.A., et al.~1997, \apj, 491, 436
%\bibitem[An et al.(2007)]{an07} An, D., Terndrup, D.~M., Pinsonneault, M.~H., et al.\ 2007, \apj, 655, 233 
%\bibitem[Anderson \& King (2000)]{andking00} Anderson, J.~\& King, I.~R.\ 2000, \pasp, 112, 1360
%\bibitem[Anderson \& King (2004)]{andking04} Anderson, J.~\& King, I.~R.\ 2004, Hubble Space Telescope
%   Advanced Camera for Surveys Instrument Science Report 04-15
%\bibitem[Bachelet et al.(2012)]{bachelet12} Bachelet, E., Fouqu{\'e}, P., Han, C., et al.\ 2012, \aap, 547, A55
%\bibitem[Baraffe et al.(2005)]{baraffe05} Baraffe, I., et al.\ 2005, \aap, 436, L47
%\bibitem[Barry(2010)]{barry_psf} Barry, R.K., et al.\ 2010,  Proc.\ SPIE, 7731, 77313
%\bibitem[Batalha et al.(2013)]{kepler_16mon} Batalha, N.~M., Rowe, J.~F., Bryson, S.~T., et al.\ 2013, \apjs, 204, 24
%\bibitem[Batista et al.(2011)]{batista11} Batista, V., Gould, A., Dieters, S., et al.\ 2011, \aap, 529, A102
%\bibitem[Batista et al.(2014)]{batista14} Batista, V., Beaulieu, J.-P., Gould, A., et al.\ 2014, \apj, 780, 54 
\bibitem[Batista et al.(2015)]{batista15} Batista, V., Beaulieu, J.-P., Bennett, D.P., et al.\ 2015, \apj, 808, 170
\bibitem[Beaulieu et al.(2018)]{beaulieu18} Beaulieu, J.-P., Batista, V., Bennett, D.~P., et al.\ 2018, \aj, 155, 78
%\bibitem[Beaulieu et al.(2006)]{ogle390} Beaulieu, J.-P., Bennett, D.~P., Fouqu{\'e}, P., et al.\ 2006, \nat, 439, 437
\bibitem[Beaulieu et al.(2016)]{beaulieu16} Beaulieu, J.-P., Bennett, D.~P., Batista, V., et al.\ 2016, \apj, 824, 83 
%\bibitem[Beichman et al.(2013)]{beichman13} Beichman, C., Gelino, C.~R., Kirkpatrick, J.~D., et al.\ 2013, \apj, 764, 101 
%\bibitem[Benedict et al.(2016)]{benedict16} Benedict, G.~F., Henry, T.~J., Franz, O.~G., et al.\ 2016, \aj, 152, 141
\bibitem[Bennett(2008)]{bennett_rev} Bennett, D.P, 2008, in Exoplanets, 
   Edited by John Mason.~Berlin: Springer.~ ISBN: 978-3-540-74007-0,  (arXiv:0902.1761)
\bibitem[Bennett(2010)]{bennett-himag} Bennett, D.P.\ 2010, \apj, 716, 1408
%\bibitem[Bennett et al.(1993)]{bennett-sod} Bennett, D.~P., Alcock, C., Allsman, R., et al.\ 1993, Bulletin of the American Astronomical Society, 25, 1402
\bibitem[Bennett et al.(2018a)]{bennett18_wfirst} Bennett, D.~P., Akeson, R., Anderson, J., et al.\ 2018a, (arXiv:1803.08564)
%\bibitem[Bennett et al.(2010a)]{bennett_MPF} Bennett, D.~P., Anderson, J., Beaulieu, J.-P., et al.\ 2010a, arXiv:1012.4486
%\bibitem[Bennett et al.(2006)]{bennett06} Bennett, D.~P., Anderson, J., Bond, I.~A., Udalski, A., \& Gould, A.\ 2006, \apjl, 647, L171
%\bibitem[Bennett et al.(2007)]{bennett07} Bennett, D.P., Anderson, J., \& Gaudi, B.S.\ 2007, \apj, 660, 781
\bibitem[Bennett et al.(2014)]{bennett14} Bennett, D.~P., Batista, V., Bond, I.~A., et~al.\ 2014, \apj, 785, 155
%\bibitem[Bennett et al.(2015)]{bennett15} Bennett, D.~P., Bhattacharya, A., Anderson, J., et al.\ 2015, \apj, 808, 169
\bibitem[Bennett et al.(2020)]{bennett20} Bennett, D.~P., Bhattacharya, A., Beaulieu, J.-P., et al.\ 2020, \aj, 159, 68
%\bibitem[Bennett et al.(2008)]{bennett08}Bennett, D.~P., Bond, I.~A., Udalski, A., et al.\ 2008, \apj, 684, 663
\bibitem[Bennett \& Rhie(1996)]{bennett96}Bennett, D.P. \& Rhie, S.H.\ 1996, \apj, 472, 660
\bibitem[Bennett \& Rhie(2002)]{bennett02}Bennett, D.P. \& Rhie, S.H.\ 2002, \apj, 574, 985
\bibitem[Bennett et al.(2010)]{bennett-ogle109} Bennett, D.~P., Rhie, S.~H., Nikolaev, S., et~al.\ 2010, \apj, 713, 837
\bibitem[Bennett et al.(2016)]{bennett16}Bennett, D.P., Rhie, S.H., Udalski, A., et al.\ 2016, \aj, 152, 125
%\bibitem[Bennett et al.(2012)]{bennett12} Bennett, D.~P., Sumi, T., Bond, I.~A., et al.\ 2012, \apj, 757, 119 
%\bibitem[Bennett et al.(2018b)]{bennett18} Bennett, D.~P., Udalski, A., Han, C., et al.\ 2018b, \aj, 155, 141 
%\bibitem[Bensby et al.(2011)]{bensby11} Bensby, T., Ad{\'e}n, D., Mel{\'e}ndez, J., et al.\ 2011, \aap, 533, A134
%\bibitem[Bhattacharya et al.(2017)]{aparna17} Bhattacharya, A., Bennett, D.~P., Anderson, J., et al.\ 2017, \aj, 154, 59 
\bibitem[Bhattacharya et al.(2018)]{aparna18} Bhattacharya, A., Beaulieu, J.-P., Bennett, D.~P., et al.\ 2018, \aj, 156, 289 
\bibitem[Bhattacharya et al.(2020)]{aparna20} Bhattacharya, A., et al.\ 2020, in preparation
\bibitem[Bond et al.(2001)]{bond01} Bond, I.~A., Abe, F., Dodd, R.~J., et al.\ 2001, \mnras, 327, 868
\bibitem[Bond et al.(2017)]{bond17} Bond, I.~A., Bennett, D.~P., Sumi, T., et al.\ 2017, \mnras, 469, 2434 
%\bibitem[Bond et~al.(2004)]{bond04} Bond, I.~A., Udalski, A., Jaroszy{\'n}ski, M.\ 2004,  \apjl, 606, L155
%\bibitem[Bonfils et al.(2011)]{bonfils11} Bonfils, X., Delfosse, X., Udry, S., et al.\ 2011, arXiv:1111.5019
%\bibitem[Borucki et al.(2011)]{borucki11} Borucki, W.~J., Koch, D.~G., Basri, G., et al.\ 2011, \apj, 736, 19
%\bibitem[Boss(1997)]{boss97} Boss, A.~P.\ 1997, Science, 276, 1836
%\bibitem[Boss(2006)]{boss06} Boss, A.~P.\ 2006, \apj, 643, 501
%\bibitem[Boss(2006)]{boss06} Boss, A.P.\ 2006, \apjl, 644, L79
%\bibitem[Bowler et al.(2011)]{bowler11} Bowler, B.~P., Liu, M.~C., Kraus, A.~L., Mann, A.~W., \& 
%Ireland, M.~J.\ 2011, \apj, 743, 148
\bibitem[Boyajian et al.(2014)]{boyajian14} Boyajian, T.S., van Belle, G., \& von Braun, K.,\
   2014, \aj, 147, 47
%\bibitem[Bressan et al.(2012)]{bressan12_PARSEC} Bressan, A., Marigo, P., Girardi, L., et al.\ 2012, \mnras, 427, 127 
%\bibitem[Brown et al.(2013)]{lcogt} Brown, T.~M., Baliber, N., Bianco, F.~B., et al.\ 2013, \pasp, in press (arXiv:1305.2437) 
%\bibitem[Butler et al.(2006)]{butler-catalog} Butler, R.~P., Wright, J.~T., Marcy, G.~W., et al.\ 2006, \apj, 646, 505
%\bibitem[Burke et al.(2015)]{burke15} Burke, C.~J., Christiansen, J.~L., Mullally, F., et al.\ 2015, \apj, 809, 8 
%\bibitem[Bramich(2008)]{bramich08} Bramich, D.M.\ 2008, \mnras, 386, L77
%\bibitem[Cardelli et al.(1989)]{cardelli89}  Cardelli, J.A., Clayton, G.C., \& Mathis, J.S.\ 1989, \apj, 345, 245
%\bibitem[Cassan et al.(2012)]{cassan12}Cassan, A., Kubas, D., Beaulieu, J.-P., et al.\ 2012,  \nat, 481, 167
%\bibitem[Carpenter(2001)]{2mass_cal} Carpenter, J.M.\ 2001, \aj 121, 2851
%\bibitem[Claret(2000)]{claret00} Claret, A.\ 2000, \aap, 363, 1081
%\bibitem[Chang \& Refsdal(1979)]{chang-refsdal79} Chang, K., \& Refsdal, S.\ 1979, \nat, 282, 561
%\bibitem[Chang \& Refsdal(1984)]{chang-refsdal84} Chang, K., \& Refsdal, S.\ 1979, \aap, 132, 168
\bibitem[Chauvin et al.(2011)]{chauvin11} Chauvin, G., Beust, H., Lagrange, A.-M., et al.\ 2011, \aap, 528, A8
%\bibitem[Chen et al.(2015)]{chen15_PARSEC} Chen, Y., Bressan, A., Girardi, L., et al.\ 2015, \mnras, 452, 1068 
%\bibitem[Chen et al.(2014)]{chen14_PARSEC} Chen, Y., Girardi, L., Bressan, A., et al.\ 2014, \mnras, 444, 2525 
%\bibitem[Christiansen et al.(2011)]{epoxi} Christiansen, J.L., et al.\ 2011, \apj, 726, 94
%\bibitem[Clarkson et al.(2008)]{clarkson08} Clarkson, W., Sahu, K., Anderson, J., et al.\ 2008, \apj, 684, 1110-1142 
\bibitem[Correia et al.(2008)]{correia08} Correia, A.~C.~M., Udry, S., Mayor, M., et al.\ 2008, \aap, 479, 271
%\bibitem[Cumming et al.(2008)]{cumming08}Cumming, A., Butler, R.~P., Marcy, G.~W., Vogt, S.~S., Wright, J.~T., \& Fischer, D.~A.\ 2008, \pasp, 120, 531
\bibitem[Dan{\v{e}}k \& Heyrovsk{\'y}(2015)]{danek15} Dan{\v{e}}k, K., \& Heyrovsk{\'y}, D.\ 2015, \apj, 806, 99
\bibitem[Dan{\v{e}}k \& Heyrovsk{\'y}(2019)]{danek19} Dan{\v{e}}k, K., \& Heyrovsk{\'y}, D.\ 2019, \apj, 880, 72
%\bibitem[D'Angelo et al.(2010)]{dangelo_book} D'Angelo, G., Durisen, R.~H., \& Lissauer, J.~J.\ 2010,
%   in Exoplanets, ed.\ S.\ Seager (Tucson, AZ: Univ. Arizona Press), 319
%\bibitem[Delfosse et al.(2000)]{delfosse00} Delfosse, X., Forveille, T., S{\'e}gransan, D., et al.\ 2000, \aap, 364, 217
%\bibitem[Delorme et al.(2012)]{delhome12} Delorme, P., Gagn{\'e}, J., Malo, L., et al.\ 2012, \aap, 548, A26 
%\bibitem[Desidera \& Barbieri(2007)]{desidera07} Desidera, S., \& Barbieri, M.\ 2007, \aap, 462, 345
%\bibitem[Di Stefano \& Esin(1995)]{distefano95} Di Stefano, R., \& Esin, A.~A.\ 1995, \apjl, 448, L1
%\bibitem[Di Stefano \& Scalzo(1999)]{distefano99} Di Stefano, R., \& Scalzo, R.~A.\ 1999, \apj, 512, 579
%\bibitem[Dolphin(2000)]{hstphot} Dolphin, A.~E.\ 2000, \pasp, 112, 1383 
%\bibitem[Dong et al.(2006)]{dong06} Dong, S., et al.\ 2006, , \apj, 642, 842
%\bibitem[Dong et al.(2007)]{dong-ogle05smc1} Dong, S., et al.\ 2007, \apj, 664, 862
%\bibitem[Dong et al.(2009)]{dong-ogle71} Dong, S., Bond, I.~A., Gould, A., et al.\ 2009, \apj, 698, 1826
%\bibitem[Dong et al.(2009b)]{dong-moa400} Dong, S., Bond, I.~A., Gould, A., et al.\ 2009, \apj, 698, 1826
%\bibitem[Doolin \& Blundell(2011)]{doolin11} Doolin, S., \& Blundell, K.~M.\ 2011, \mnras, 418, 2656
\bibitem[Doyle et al.(2011)]{doyle11} Doyle, L.~R., Carter, J.~A., Fabrycky, D.~C., et al.\ 2011, Science, 333, 1602 
%\bibitem[Drimmel \& Spergel(2001)]{drimmel}  Drimmel, R., \& Spergel, D.~N.\ 2001, \apj, 556, 181
%\bibitem[Dumusque et al.(2012)]{alpha_cen_Bb12} Dumusque, X., Pepe, F., Lovis, C., et al.\ 2012, \nat, 491, 207
%\bibitem[Fruchter \& Hook(2002)]{drizzle} Fruchter A.S.\ 2002, \pasp, 114, 144
%\bibitem[Ford \& Rasio(2008)]{ford08} Ford, E.~B., \& Rasio, F.~A.\ 2008, \apj, 686, 621
%\bibitem[Fukui et al.(2015)]{fukui15} Fukui, A., Gould, A., Sumi, T., et al.\ 2015, \apj, 809, 74 
%\bibitem[Furusawa et al.(2013)]{moa328} Furusawa, K., Udalski, A., Sumi, T., et al.\ 2013, \apj, 779, 91
%\bibitem[Gaudi(2010)]{gaudi_rev}Gaudi, B.S.\ 2010, in Exoplanets, ed. S. Seager (Tucson: University of 
%Arizona Press), 79 (arXiv:1002.0332)
\bibitem[Gaudi(2012)]{gaudi_araa} Gaudi, B.~S.\ 2012, \araa, 50, 411
\bibitem[Gaudi et al.(2008)]{gaudi-ogle109} Gaudi, B.~S., Bennett, D.~P., Udalski, A., et al.\ 2008, Science, 319, 927
%\bibitem[Gaudi \& Gould(1997)]{gaudi97} Gaudi, B.S., \& Gould, A.\ 1997, \apj, 486, 85
%\bibitem[Gonzalez et al.(2011)]{vvv_extinct} Gonzalez, O.~A., Rejkuba, M., Zoccali, M., Valenti, E., \& Minniti, D.\ 2011, \aap, 534, A3
%\bibitem[Gould(1992)]{gould-par1} Gould, A.\ 1992, \apj, 392, 442
%\bibitem[Gould(2004)]{gould-jerk} Gould, A.\ 2004, \apj, 606, 319
%\bibitem[Gould(2008)]{gould-hex} Gould, A.\ 2008, \apj, 681, 1593
%\bibitem[Gould(2014)]{gould-1dpar} Gould, A.\ 2014, J.\ Kor.\ Ast.\ Soc., 47, 215
%\bibitem[Gould et al.(2010)]{gould_col} Gould, A., Dong, S., Bennett, D.~P., et al.\ 2010a, \apj, 710, 1800
%\bibitem[Gould et al.(2010b)]{gould10}Gould, A., Dong, S., Gaudi, B.S., et al.\ 2010b,  \apj, 720, 1073
%\bibitem[Gould et al.(2004)]{gould04}Gould, A., Gaudi, B.S., \& Han, C., 2004, arXiv:astro-ph/0405217
\bibitem[Gould \& Loeb(1992)]{gouldloeb92} Gould, A. \& Loeb, A. 1992, \apj, 396, 104
%\bibitem[Gould et al.(2006)]{gould06} Gould, A., Udalski, A., An, D., et al.\ 2006, \apjl, 644, L37
%\bibitem[Gould et al.(2009)]{gould09} Gould, A., Udalski, A., Monard, B., et al.\ 2009, \apjl, 698, L147
\bibitem[Gould et al.(2014)]{gould14} Gould, A., Udalski, A., Shin, I.-G., et al.\ 2014, Science, 345, 46 
%\bibitem[Gould \& Yee(2013)]{gould_yee_terpar13} Gould, A., \& Yee, J.~C.\ 2013, \apj, 764, 107
%\bibitem[Green et al.(2012)]{WFIRST_rep} Green, J., Schechter, P., Baltay, C., et al.\ 2012, arXiv:1208.4012 
%\bibitem[Griest \& Hu(1992)]{griest92} Griest, K., \& Hu, W.\ 1992, \apj, 397, 362
%\bibitem[Griest \& Safizadeh(1998)]{griest98} Griest, K., \& Safizadeh, N.\ 1998, \apj, 500, 37
%\bibitem[Guillochon et al.(2011)]{guillochon11} Guillochon, J., Ramirez-Ruiz, E., \& Lin, D.\ 2011, \apj, 732, 74 
%\bibitem[Han et al.(2016)]{han_ob130723} Han, C., Bennett, D.~P., Udalski, A., \& Jung, Y.~K.\ 2016, \apj, in press (arXiv:1604.06533)
%\bibitem[Han \& Gould(1997)]{han97} Han , C., \& Gould, A.\ 1997, \apj, 480, 196
%\bibitem[Han et al.(2005)]{han_widepl2005} Han, C., Gaudi, B.~S., An, J.~H., \& Gould, A.\ 2005, \apj, 618, 962 
%\bibitem[Han \& Kang(2003)]{han_widepl2003} Han, C., \& Kang, Y.~W.\ 2003, \apj, 596, 1320
\bibitem[Han et al.(2017)]{han_ob160613} Han, C., Udalski, A., Gould, A., et al.\ 2017, \aj, 154, 223
%\bibitem[Hartman et al.(2004)]{hartmanISIS} Hartman, J. D., Bakos, G., Stanek, K. Z., \& Noyes, R. W. 2004, AJ, 128, 1761
\bibitem[Hatzes et al.(2003)]{hatzes03} Hatzes, A.~P., Cochran, W.~D., Endl, M., et al.\ 2003, \apj, 599, 1383
%\bibitem[Henderson et al.(2014)]{henderson14} Henderson, C.~B., Park, H., Sumi, T., et al.\ 2014, \apj, 794, 71
%\bibitem[Henry et al.(1999)]{henry99} Henry, T.~J., Franz, O.~G., Wasserman, L.~H., et al.\ 1999, \apj, 512, 864 
%\bibitem[Henry \& McCarthy(1993)]{henry93} Henry, T.~J., \& McCarthy, D.~W., Jr.\ 1993, \aj, 106, 773
%\bibitem[Hirao et al.(2017)]{hirao17} Hirao, Y., Udalski, A., Sumi, T., et al.\ 2017, \aj, 154, 1 
%\bibitem[Hilton(2011)]{hilton11} Hilton, E.~J.\ 2011, PhD Thesis, University of Washington
%\bibitem[Hilton et al.(2010)]{hilton10} Hilton, E.~J., Hawley, S.~L., Kowalski, A.~F., \& Holtzman, J.\ 2010, arXiv:1012.0577 
\bibitem[Holman \& Wiegert(1999)]{holman99} Holman, M.~J., \& Wiegert, P.~A.\ 1999, \aj, 117, 621 
%\bibitem[Heyrovsky(2003)]{heyrovsky03} Heyrovsk\'y, D.\ 2003, \apj, 594, 464
%\bibitem[Heyrovsky(2007)]{heyrovsky07} Heyrovsk\'y, D.\ 2007, \apj, 656, 483
\bibitem[Holtzman et al.(1998)]{holtzman98} Holtzman, J.~A., Watson, A.~M., Baum, W.~A., et al.\ 1998, \aj, 115, 1946 
%\bibitem[Howard et al.(2010)]{howard10}Howard, A.W. et al.\ 2010, Science, 330, 653
%\bibitem[Hubickyj et al.(2005)]{hubickyj05} Hubickyj, O., Bodenheimer, P., \& Lissauer, J.J.\ 2005, Icarus, 179, 415
%\bibitem[Hwang et al.(2018)]{hwang18} Hwang, K.-H., Udalski, A., Shvartzvald, Y., et al.\ 2018, \aj, 155, 20
\bibitem[Hwang et al.(2019)]{hwang19} Hwang, K.-H., Ryu, Y.-H., Kim, H.-W., et al.\ 2019, \aj, 157, 23
\bibitem[Ida \& Lin(2004)]{idalin04} Ida, S.\ \& Lin, D.N.C.\ 2004, \apj, 604, 388
%\bibitem[Ida \& Lin(2005)]{ida05} Ida, S., \& Lin, D.N.C.\ 2005, \apj, 626, 1045
%\bibitem[Janczak et al.(2010)]{janczak10} Janczak, J., Fukui, A., Dong, S., et al.\ 2010, \apj, 711, 731
\bibitem[Jaroszy{\'n}ski et al.(2010)]{jar10} Jaroszy{\'n}ski, M., Skowron, J., Udalski, A., et al.\ 2010, \actaa, 60, 197
%\bibitem[Johnson et al.(2007)]{johnson07} Johnson, J.~A., Butler, R.~P., Marcy, G.~W., et al.\ 2007, \apj, 670, 833
%\bibitem[Johnson et al.(2010)]{johnson10} Johnson, J.~A., Aller, K.~M., Howard, A.~W., \& Crepp, J.~R.\ 2010, \pasp, 122, 905 
\bibitem[Kaib et al.(2013)]{kaib13} Kaib, N.~A., Raymond, S.~N., \& Duncan, M.\ 2013, \nat, 493, 381 
%\bibitem[Kennedy \& Kenyon(2008)]{kennedy_snowline} Kennedy, G.~M., \& Kenyon, S.~J.\ 2008, \apj, 673, 502 
%\bibitem[Kennedy et al.(2006)]{kennedy-searth} Kennedy, G.M., Kenyon, S.J.,  \& Bromley, B.C.\ 2006, \apjl 650, L139
%\bibitem[Kenyon \& Hartmann(1995)]{kenyon95} Kenyon, S.~J., \& Hartmann, L.\ 1995, \apjs, 101, 117 
\bibitem[Kervella et al.(2004)]{kervella_dwarf} Kervella, P., Th{\'e}venin, F., Di Folco, E., \& S{\'e}gransan, D.\ 2004, \aap, 426, 297
%\bibitem[Kervella et al.(2004)]{kervella04g} Kervella P., et al.\ 2004,  \aap, 428, 587
%\bibitem[Kervella \& Fouqu{\'e}(2008)]{kervella08} Kervella, P., \& Fouqu{\'e}, P.\ 2008, \aap, 491, 855
%\bibitem[Kim et al.(2013)]{kmtnet_pipe} Kim, D.~J., Lee, C.~U., Kim, S.~L., \& Park, B.~G.\ 2013, Pub.\ Korean Astron.\ Soc., 28, 1
%\bibitem[Kim et al.(2010)]{kmtnet10} Kim, S.-L., Park, B.-G., Lee, C.-U., et al.\ 2010, \procspie, 7733, 77733
%\bibitem[Kim et al.(2016)]{kmtnet} Kim, S.-L., Lee, C.-U., Park, B.-G., et al.\ 2016, Journal of Korean Astronomical Society, 49, 37 
\bibitem[Kondo et al.(2019)]{kondo19} Kondo, I., Sumi, T., Bennett, D.~P., et al.\ 2019, \aj, 158, 224
%\bibitem[Koshimoto et al.(2017a)]{kosh17_ob120950} Koshimoto, N., Udalski, A., Beaulieu, J.~P., et al.\ 2017a, \aj, 153, 1 
%\bibitem[Koshimoto et al.(2017b)]{kosh17_mb16227} Koshimoto, N., Shvartzvald, Y., Bennett, D.~P., et al.\ 2017b, \aj, 154, 3
\bibitem[Koshimoto et al.(2020)]{koshimoto20} Koshimoto, N., Bennett, D.~P., \& Suzuki, D.\ 2019, AJ, in press (arXiv:1910.11448)
\bibitem[Koshimoto et al.(2014)]{koshimoto14} Koshimoto, N., Udalski, A., Sumi, T., et al.\ 2014, \apj, 788, 128
\bibitem[Kostov et al.(2014)]{kostov14} Kostov, V.~B., McCullough, P.~R., Carter, J.~A., et al.\ 2014, \apj, 784, 14 
\bibitem[Kostov et al.(2013)]{kostov13} Kostov, V.~B., McCullough, P.~R., Hinse, T.~C., et al.\ 2013, \apj, 770, 52 
\bibitem[Kostov et al.(2016)]{kostov16} Kostov, V.~B., Orosz, J.~A., Welsh, W.~F., et al.\ 2016, \apj, 827, 86 
%\bibitem[Kowalski et al.(2009)]{kowalski09} Kowalski, A.~F., Hawley, S.~L., Hilton, E.~J., et al.\ 2009, \aj, 138, 633
%\bibitem[Kowalski et al.(2010)]{kowalski10} Kowalski, A.~F., Hawley, S.~L., Holtzman, J.~A., Wisniewski, J.~P., \& Hilton, E.~J.\ 2010, \apjl, 714, L98
%\bibitem[Koz{\l}owski et al.(2006)]{koz06} Koz{\l}owski, S., Wo{\'z}niak, P.~R., Mao, S., et al.\ 2006, \mnras, 370, 435
\bibitem[Kraus et al.(2016)]{kraus16} Kraus, A.~L., Ireland, M.~J., Huber, D., et al.\ 2016, \aj, 152, 8
%\bibitem[Kubas et al.(2012)]{moa192_naco} Kubas, D., Beaulieu, J.~P., Bennett, D.~P., et al.\ 2012, \aap, 540, A78
%\bibitem[Kurucz(1993a)]{kurucz93a} Kurucz, R.L.\ 1993a, Kurucz CD-ROM 16, (SAO, Cambridge, MA, 1993).
%\bibitem[Kurucz(1993b)]{kurucz93b} Kurucz, R.L.\ 1993b, Kurucz CD-ROM 17,  (SAO, Cambridge, MA, 1993)
%\bibitem[Kurucz(1994)]{kurucz94} Kurucz, R.L.\ 1994, Kurucz CD-ROM 19,  (SAO, Cambridge, MA, 1993)
%\bibitem[Kurucz(1996)]{kurucz} Kurucz, R.L.\ 1996, ASP Conference Series, 108, 2
%\bibitem[Lagrange et al.(2006)]{lagrange06} Lagrange, A.-M., Beust, H., Udry, S., et al.\ 2006, \aap, 459, 955
%\bibitem[Laughlin et al.(2004)]{laughlin04} Laughlin, G.  Bodenheimer, P.\ \& Adams, F.C.\ 2004, \apjl, 612, L73
%\bibitem[Lecar(2006)]{lecar_snowline} Lecar, M., Podolak, M., Sasselov, D., \& Chiang, E.\ 2006, \apj, 640, 1115
%\bibitem[Levison et~al.(1998)]{levison98} Levison, H.~F., Lissauer, J.~J., Duncan, M.~J.\ 1998, \aj, 116, 1998
\bibitem[Lissauer(1993)]{lissauer_araa} Lissauer, J.J.\ 1993, Ann.\ Rev.\ Astron.\ Ast., 31, 129
%\bibitem[Malmberg et al.(2011)]{malmberg11} Malmberg, D., Davies, M.~B., \& Heggie, D.~C.\ 2011, \mnras, 411, 859
%\bibitem[Lu et al.(2014)]{lu14} Lu, J.~R., Neichel, B., Anderson, J., et al.\ 2014, \procspie, 9148, 91480B 
%\bibitem[Lu et al.(2016)]{lu16} Lu, J.~R., Sinukoff, E., Ofek, E.~O., Udalski, A., \& Kozlowski, S.\ 2016, \apj, 830, 41 
%\bibitem[Mao \& Paczy\'{n}ski(1991)]{mao91}Mao, S., \& Paczy\'{n}ski, B.\ 1991, \apj, 374, L37
\bibitem[Marzari \& Thebault(2019)]{marzari19} Marzari, F., \& Thebault, P.\ 2019, Galaxies, 7, 84
%\bibitem[Mayor \& Queloz(2012)]{mayor12} Mayor, M., \& Queloz, D.\ 2012, \nar, 56, 19
%\bibitem[Minniti et al.(2010)]{minniti-vvv} Minniti, D., Lucas, P.~W., Emerson, J.~P., et al.\ 2010, \na, 15, 433 
%\bibitem[Musielak et al.(2005)]{musielak05} Musielak, Z.~E., Cuntz, M., Marshall, E.~A., \& Stuit, T.~D.\ 2005, \aap, 434, 355 
%\bibitem[Miyake et al.(2011)]{miyake11} Miyake, N., Sumi, T., Dong, S., et al.\ 2011, \apj, 728, 120
%\bibitem[Montet et al.(2013)]{montet13} Montet, B.T., Crepp, J.R., Johnson, J.A., Howard, A.W., \& Marcy, G.W.\ 2013, \apj, submitted (arXiv:1307.5849 )
\bibitem[Mordasini et al.(2009)]{mor09} Mordasini, C., Alibert, Y., \& Benz, W.\ 2009, \aap, 501, 1139 
%\bibitem[Movshovitz \& Podolak(2008)]{movshovitz08} Movshovitz, N., \& Podolak, M.\ 2008, Icarus, 194, 368
%\bibitem[Mr{\'o}z et al.(2017a)]{mroz17_ob160596} Mr{\'o}z, P., Han, C., Udalski, A., et al.\ 2017a, \aj, 153, 143 
%\bibitem[Mr{\'o}z et al.(2017b)]{mroz17_2pl} Mr{\'o}z, P., Udalski, A., Bond, I.~A., et al.\ 2017b, \aj, 154, 205 
\bibitem[Mugrauer \& Neuh{\"a}user(2009)]{mugrauer09} Mugrauer, M., \& Neuh{\"a}user, R.\ 2009, \aap, 494, 373
%\bibitem[Mullally et al.(2016)]{mullaly16} Mullally, F., Coughlin, J.~L., Thompson, S.~E., et al.\ 2016, arXiv:1602.03204 
%\bibitem[Muraki et al.(2011)]{muraki11} Muraki, Y., Han, C., Bennett, D.~P., et al.\ 2011, \apj, 741, 22
%\bibitem[Nagasawa \& Ida(2011)]{nagasawa11} Nagasawa, M., \& Ida, S.\ 2011, \apj, 742, 72 
\bibitem[Nataf et al.(2013)]{nataf13} Nataf, D.~M., Gould, A., Fouqu{\'e}, P., et al.\ 2013, \apj, 769, 88 
\bibitem[Nayakshin et al.(2019)]{nayaskshin19} Nayakshin, S., Dipierro, G., \& Szul{\'a}gyi, J.\ 2019, \mnras, 488, L12
\bibitem[Nelson(2000)]{nelson00} Nelson, A.~F.\ 2000, \apjl, 537, L65
\bibitem[Neuh{\"a}user et al.(2007)]{neuhauser07} Neuh{\"a}user, R., Mugrauer, M., Fukagawa, M., et al.\ 2007, \aap, 462, 777
%\bibitem[Nishiyama et al.(2009)]{nish09} Nishiyama, S., Tamura, M., Hatano, H., et al.\ 2009, \apj, 696, 1407
\bibitem[Orosz et al.(2012)]{orosz12} Orosz, J.~A., Welsh, W.~F., Carter, J.~A., et al.\ 2012, Science, 337, 1511 
\bibitem[Paardekooper et al.(2008)]{paar08}Paardekooper, S.-J., Thebault, P., \& Mellema, G., 2008, MNRAS, 386, 973
%\bibitem[Park et al.(2012)]{kmtnet} Park, B.-G., Kim, S.-L., Lee, J.~W., et al.\ 2012, \procspie, 8444, 844447-1
%\bibitem[Pejcha \& Heyrovsky(2009)]{pei_hey} Pejcha, O., \& Heyrovsk\'y, D.\ 2009, \apj, 690, 1772
%\bibitem[Penny et al.(2013)]{penny13} Penny, M.~T., Kerins, E., Rattenbury, N., et al.\ 2013,  \mnras, submitted (arXiv:1206.5296)
\bibitem[Penny et al.(2019)]{penny19} Penny, M.~T., Gaudi, B.~S., Kerins, E., et al.\ 2019, \apjs, 241, 3
\bibitem[Picogna \& Marzari(2013)]{picogna13} Picogna, G., \& Marzari, F.\ 2013, \aap, 556, A148
%\bibitem[Pietrukowicz et al.(2011)]{m22ml} Pietrukowicz, P., Minniti, D., Jetzer, P., Alonso-Garcia, J., \& Udalski, A.\ 2011, \apj, 744, L18
%\bibitem[Poindexter et al.(2005)]{poindexter05} Poindexter, S., et al.\ 2005, \apj, 633, 914
\bibitem[Poleski et al. (2014)]{ogle092} Poleski R., Skowron J., Udalski A., et al.,\ 2014, ApJ, 795, 1
\bibitem[Pollack et al.(1996)]{pollack96} Pollack, J.~B., Hubickyj, O., Bodenheimer, P., et al.\ 1996, \icarus, 124, 62
%\bibitem[Popowski et al.(2003)]{pop_extinct} Popowski, P., Cook, K.~H., \& Becker, A.~C.\ 2003, \aj, 126, 2910
%\bibitem[Quanz et al.(2012)]{quanz12} Quanz, S.~P., Lafreni{\`e}re, D., Meyer, M.~R., Reggiani, M.~M., \& Buenzli, E.\ 2012, \aap, 541, A133 
%\bibitem[Queloz et al.(2000)]{queloz00} Queloz, D., Mayor, M., Weber, L., et al.\ 2000, \aap, 354, 99
%\bibitem[Ranc et al.(2015)]{ranc15} Ranc, C., Cassan, A., Albrow, M.~D., et al.\ 2015, \aap, 580, A125
%\bibitem[Ranc et al.(2018)]{ranc18} Ranc, C., et al.\ 2018, in preparation
%\bibitem[Rafikov(2011)]{rafikov11} Rafikov, R.\ 2011, \apj, 727, 86
%\bibitem[Rattenbury et al.(2007)]{rattenbury07} Rattenbury, N.~J., Mao, S., Debattista, V.~P., et al.\ 2007, \mnras, 378, 1165
%\bibitem[Rattenbury et al.(2015)]{rattenbury15} Rattenbury, N.~J., Bennett, D.~P., Sumi, T., et al.\ 2015, \mnras, 454, 946 
%\bibitem[Rattenbury et al.(2017)]{rattenbury17} Rattenbury, N.~J., Bennett, D.~P., Sumi, T., et al.\ 2017, \mnras, 466, 2710 
%\bibitem[Refsdal(1966)]{refsdal-par} Refsdal, S. 1966, \mnras, 134, 315
\bibitem[Rhie et al.(1999)]{rhie_98smc1} Rhie, S.~H., Becker, A.~C., Bennett, D.~P., et al.\ 1999, \apj, 522, 1037
%\bibitem[Rhie et al.(2000)]{rhie00} Rhie, S.~H., Bennett, D.~P., Becker, A.~C., et al.\ 2000, \apj, 533, 378
%\bibitem[Robin et al.(2003)]{robin03} Robin, A.~C., Reyl{\'e}, C., Derri{\`e}re, S., \& Picaud, S.\ 2003, \aap, 409, 523
\bibitem[Roell et al.(2012)]{roell12} Roell, T., Neuh{\"a}user, R., Seifahrt, A., et al.\ 2012, \aap, 542, A92
%\bibitem[Sako et al.(2008)]{sako_moacam3} Sako, T., Sekiguchi, T., Sasaki, M., et al.\ 2008, Experimental Astronomy, 22, 51
%\bibitem[Schechter, Mateo, \& Saha(1993)]{dophot} Schechter, P.~L., Mateo, M., \& Saha, A.\ 1993, \pasp, 105, 1342
%\bibitem[Shvartzvald et al.(2018)]{shvartzvald18} Shvartzvald, Y., Calchi Novati, S., Gaudi, B.~S., et al.\ 2018, \apjl, 857, L8 
%\bibitem[Shvartzvald et al.(2016)]{shvartzvald16} Shvartzvald, Y., Maoz, D., Udalski, A., et al.\ 2016, \mnras, 457, 4089 
%\bibitem[Skowron(2011)]{skowron11} Skowron, J., et al.\ 2011, \apj, submitted, (arXiv:1101.3312)
%\bibitem[Skowron et al.(2013)]{skowron13} Skowron, J., et al.\ 2011, \apj, submitted
%\bibitem[Smith et al.(2002)]{smith02} Smith, M.C., Mao, S., \& Wo\'zniak, P.\ 2002, \mnras, 332, 962
%\bibitem[Smith et al.(2003)]{smith03} Smith, M.C., Mao, S., \& Paczy{\'n}ski, B.\ 2003, \mnras, 339, 925
\bibitem[Smullen et al.(2016)]{smullen_kratter16} Smullen, R.~A., Kratter, K.~M., \& Shannon, A.\ 2016, \mnras, 461, 1288
\bibitem[Spergel et al.(2015)]{WFIRST_AFTA} Spergel, D., Gehrels, N., Baltay, C., et al.\ 2015, arXiv:1503.03757 
%\bibitem[Stetson(1994)]{allframe} Stetson, P.B.\ 1994, \pasp, 106, 250
%\bibitem[Street et al.(2016)]{street16} Street, R.~A., Udalski, A., Calchi Novati, S., et al.\ 2016, \apj, 819, 93
%\bibitem[Stubbs et al.(2007)]{stubbs07} Stubbs, C.W., et al.\ 2007, \pasp, 119, 1163
%\bibitem[Sumi et al.(2010)]{sumi10}Sumi, T., Bennett, D.~P., Bond, I.~A. et al.\ 2010,  \apj, 710, 1641
%\bibitem[Sumi et al.(2011)]{sumi11}Sumi, T., Kamiya, K., Bennett, D.~P., et al.\ 2011,  \nat, 473, 349
%\bibitem[Sumi et al.(2016)]{sumi16} Sumi, T., Udalski, A., Bennett, D.~P., et al.\ 2016, \apj, 825, 112 
%\bibitem[Sumi et al.(2004)]{sumi04} Sumi, T., Wu, X., Udalski, A., et al.\ 2004, \mnras, 348, 1439
\bibitem[Suzuki et al.(2018)]{suzuki18} Suzuki, D., Bennett, D.~P., Ida, S., et al.\ 2018, \apjl, 869, L34
\bibitem[Suzuki et al.(2016)]{suzuki16} Suzuki, D., Bennett, D.~P., Sumi, T., et al.\ 2016, \apj, 833, 145
%\bibitem[Suzuki et al.(2014)]{suzuki14} Suzuki, D., Udalski, A., Sumi, T., et al.\ 2014, \apj, 780, 123 
%\bibitem[Suzuki et al.(2014e)]{suzuki14e} Suzuki, D., Udalski, A., Sumi, T., et al.\ 2014e, \apj, 788, 97 
%\bibitem[Szul{\'a}gyi et al.(2016)]{szulagyi16} Szul{\'a}gyi, J., Masset, F., Lega, E., et al.\ 2016, \mnras, 460, 2853
%\bibitem[Szul{\'a}gyi et al.(2014)]{szulagyi14} Szul{\'a}gyi, J., Morbidelli, A., Crida, A., et al.\ 2014, \apj, 782, 65
\bibitem[Szyma{\'n}ski et al.(2011)]{ogle3-phot} Szyma{\'n}ski, M.~K., Udalski, A., Soszy{\'n}ski, I., et al.\ 2011, \actaa, 61, 83 
%\bibitem[Tang et al.(2014)]{tang14_PARSEC} Tang, J., Bressan, A., Rosenfield, P., et al.\ 2014, \mnras, 445, 4287 
\bibitem[Terry et al.(2020)]{terry20} Terry, S., et al.\ 2020, in preparation
\bibitem[Thebault(2011)]{thebault11}Thebault, P., 2011, CeMDA, 111, 29
\bibitem[Thebault \& Haghighipour(2015)]{thebault_highig15} Thebault, P., \& Haghighipour, N.\ 2015, Planetary Exploration and Science: Recent Results and Advances, 309 (arXiv:1406.1357)
%\bibitem[Thommes et al.(2008)]{thommes08} Thommes, E.W., Matsumura, S., \& Rasio F.A.\ 2008, Science, 321, 814
\bibitem[Tomaney \& Crotts (1996)]{tom96}Tomaney, A.B. \& Crotts, A.P.S.\ 1996, \aj 112, 2872
%\bibitem[Traub(2011)]{traub11} Traub, W.\ 2011, ApJ, submitted.
%\bibitem[Twicken et al.(2016)]{kepler_q17} Twicken, J.~D., Jenkins, J.~M., Seader, S.~E., et al.\ 2016, ApJ, submitted (arXiv:1604.06140)
%\bibitem[Udalski(2003a)]{udalski-ext} Udalski, A.\ 2003a, \apj, 590, 284
\bibitem[Udalski(2003)]{ogle-pipeline} Udalski, A.\ 2003, \actaa, 53, 291
%\bibitem[Udalski et al.(2005)]{udalski05} Udalski, A., Jaroszy{\'n}ski, M., Paczy{\'n}ski, B., et al.\ 2005, \apjl, 628, L109
%\bibitem[Udalski et al.(2015b)]{udalski_ob130723} Udalski, A., Jung, Y.~K., Han, C., et al.\ 2015b, \apj, 812, 47 
%\bibitem[Udalski et al.(1994)]{ogle-ews} Udalski, A., Szyma\'{n}ski, M., Ka{\l}u\.{z}ny, J., Kubiak, M., Mateo, M., Krzmi\'{n}ski, W., \& \pac, B.\ 1994, Acta Astron., 44, 227
%\bibitem[Udalski et al.(2002)]{uda02} Udalski, A., Szymanski, M., Kubiak, M., et al.\ 2002, \actaa, 52, 217
\bibitem[Udalski et al.(2008)]{udalski08} Udalski, A., Szyma\'{n}ski, M., Soszy\'{n}ski, I. \& Poleski, R.\ 2008, Acta Astron., 58, 69
%\bibitem[Udalski et al.(2015a)]{ogle4} Udalski, A., Szyma{\'n}ski, M.~K., \& Szyma{\'n}ski, G.\ 2015a, \actaa, 65, 1 
\bibitem[Udalski et al.(2018)]{udalski18} Udalski, A., Ryu, Y.-H., Sajadian, S., et al.\ 2018, \actaa, 68, 1 
%\bibitem[Veras et al.(2011)]{veras11}  Veras, D., Wyatt, M.~C., Mustill, A.~J., Bonsor, A., \& Eldridge, J.~J.\ 2011, \mnras, 417, 2104
%\bibitem[Veras \& Raymond(2012)]{veras12} Veras, D., \& Raymond, S.~N.\ 2012, \mnras, 421, L117
%\bibitem[Veras \& Tout(2012)]{veras_tout12} Veras, D., \& Tout, C.~A.\ 2012, \mnras, 422, 1648 
%\bibitem[Voyatzis et al.(2013)]{voyatzis13} Voyatzis, G., Hadjidemetriou, J.~D., Veras, D., \& Varvoglis, H.\ 2013, \mnras, 430, 3383 
%\bibitem[Wambsganss(2011)]{wamb11} Wambsganss, J.\ 2011, \nat, 473, 289
%\bibitem[Ward(1997)]{ward97} Ward, W.R.\ 1997, Icarus, 126, 261
\bibitem[Welsh et al.(2012)]{welch12} Welsh, W.~F., Orosz, J.~A., Carter, J.~A., et al.\ 2012, \nat, 481, 475
\bibitem[Welsh et al.(2015)]{welch15} Welsh, W.~F., Orosz, J.~A., Short, D.~R., et al.\ 2015, \apj, 809, 26 
%\bibitem[Witt(1995)]{witt95} Witt, H.J.\ 1995, \apj, 449, 42
%\bibitem[Wright \& Gaudi(2013)]{wright_gaudi_book2013} Wright, J.~T., \& Gaudi, B.~S.\ 2013, Planets, Stars and Stellar Systems.~Volume 3: Solar and Stellar Planetary Systems, 489 
%\bibitem[Yee(2013)]{yee_WFIRST_par} Yee, J.~C.\ 2013, \apjl, 770, L31 
%\bibitem[Yee et al.(2012)]{yee12} Yee, J.~C., Shvartzvald, Y., Gal-Yam, A., et al.\ 2012, \apj, 755, 102
%\bibitem[Yee et al.(2009)]{yee09} Yee, J.~C., Udalski, A., Sumi, T., et al.\ 2009, \apj, 703, 2082 
\bibitem[Yoo et al.(2004)]{yoo_rad} Yoo, J., DePoy, D.~L., Gal-Yam, A., et al.\ 2004, \apj, 603, 139
\bibitem[Zsom et al.(2011)]{zsom11} Zsom, A., S{\'a}ndor, Z., \& Dullemond, C.~P.\ 2011, \aap, 527, A10
\bibitem[Zucker et al.(2004)]{zucker04} Zucker, S., Mazeh, T., Santos, N.~C., et al.\ 2004, \aap, 426, 695
\end{thebibliography}
\end{document}